# Localized Photon Absorption in a Single-Crystalline Material

Daniel Kazenwadel[†], Jacob Holder[†], Livio Ciorciaro, Noel Neathery,
Raphael Schwenzer, Leon Oleschko, Jannik Hertkorn, Margaretha Sandor, Peter Baum*

*Universität Konstanz, Fachbereich Physik, 78464 Konstanz, Germany*

*[†]These authors contributed equally to this work.*

**peter.baum@uni-konstanz.de*

**The absorption of light is one of the most fundamental processes in condensed-matter physics and optics. Here we investigate under which conditions laser light is absorbed by a crystalline material as an electromagnetic wave with delocalized properties or rather as photons that cause discrete, localized, nanometer-sized consequences. We excite the first-order phase transition of vanadium dioxide with laser pulses of sufficient frequency to overcome the band gap but with insufficient pulse energy to overcome the latent heat. According to Maxwell's equations and Bloch theory, no transition should occur, because nowhere in the material is enough energy. Nevertheless, we observe with ultrafast electron diffraction a disordered crystal geometry with nanometer-sized spots of switched material that grow and diminish with time. The amount of localized spots matches approximately to the number of photons in the absorbed laser wave. Two optical experiments substantiate this phenomenon, and simulations reproduce all measurements results. We discuss whether crystals defects, temperature, or a genuine wavefunction collapse can explain the discovered phenomenon. Practically, the reported absorption mechanism enables local consequences at substantially higher energy than average and provides insight into symmetry breaks and non-thermal fluctuations within complex materials.**

Light absorption is a ubiquitous and highly useful process in basic physics, material science, optics and optical technology. For example, the loss of certain optical frequencies in atoms, molecules or condensed matter can provide insight into their internal structure and dynamics through spectroscopy. Also, converting light into other forms of energy is vital for material processing, solar cells, optical sensors, or photosynthesis. Yet, the underlying physics involves an interesting fundamental enigma (*1*). According to Maxwell's equations, a laser beam is a macroscopic electromagnetic wave with non-local properties. Likewise, according to Bloch theory, the electrons in the band structure of a crystalline material are delocalized matter waves with indistinguishable local properties. However, when light is absorbed into the band structure of a material, we eventually arrive at an elevated temperature and incoherent atomic and electronic disorder as a function of spatial coordinates. For example, the carrier-hole pairs in avalanche photodiodes or solar cells are usually discussed as random, localized objects that can propagate and diffuse. Therefore, the absorption of light and the following processes somehow break the fundamental translational symmetries of Maxwell's equations and Bloch's theory (*1–3*).

Here we examine the absorption of light by a solid material with a structural phase transition and investigate whether a coherent laser beam is absorbed as a wave or whether random events create localized phase-transition spots at cost of no excitation elsewhere. These two cases are depicted in Fig. 1. If the electromagnetic wave of a laser beam (orange) is absorbed by a crystal as a whole (Fig. 1a), the deposited energy density is a wide and homogeneous spot (Fig. 1b). However, if the individual photons cause localized excitations, the deposited energy density becomes a distribution of random peaks (Fig. 1d). Depending on their diameter, these spots may have a much higher local energy density than the original laser wave and may therefore enable local consequences in the crystal that could otherwise not be achieved.

The central idea of the experiment (Fig. 2a) is the illumination of an absorbing material with a first-order phase transition and large latent heat with pulses of coherent laser light. The photon energy is above the material's band gap, such that the light can be absorbed, but we adjust the pulse energy to be too weak to reach the transition temperature and overcome the latent heat anywhere in the material. Thus, we do not generate a superheated material that could produce domains by classical nucleation theory (see Methods). Also, we ensure that the material is single-crystalline (Suppl. Fig. S1) with a low amount of defects, substantially lower than the number of absorbed photons in the laser beam (see Methods). We then use time-resolved electron diffraction and mid-infrared spectroscopy to see if nothing happens (Fig. 1a) or if parts of the material change their crystal structure in a localized way (Fig. 1c). Theoretical evidence for the possibility of such a process comes from time-dependent density functional theory in crystallographic super-cells (*4*).

The absorbing material in the experiment is a single-crystal of vanadium dioxide ($VO_2$) (*5*), a well-understood material with a first-order solid-to-solid phase transition at a temperature of ~69 °C. There, a low-temperature/monoclinic phase with vanadium dimers and insulating electric properties converts into a high-temperature/rutile-like phase with no vanadium dimers and metallic properties (*6*, *7*). The associated latent heat is rather large, ~50 kJ/kg (*6*), equivalent to ~85 meV per rutile unit cell (see Methods). The motion of the atoms from the initial to the final geometry takes only 80-300 fs (*8–12*) with no substantial dependence on initial temperature (*13*) and the associated transition from insulator to metal is also ultrafast (*7*, *8*, *14*, *15*), potentially even faster than the structural change (*10*, *16*). The vanadium atoms in the high-temperature phase have a soft potential with large phonon amplitude (*11*), transition temperatures depend on strain (*17–20*), and double-pulse excitations can provide coherent phonon control (*21*). Density functional theory of supercells with finite atomic disorder has predicted an excitation of single V-V dimers (*4*) that depends on temperature (*22*). Breaking one V-V dimer requires ~60 meV (*23*). One single laser photon at an energy of $E_{h\nu} \approx 1.2$ eV could therefore in principle transform nothing (if the laser beam remains a wave), or $E_{h\nu}/(60\ \text{meV}) \approx 20$ random unit cells (if we break random V-V bonds), or create only one transformed spot per photon (if the laser beam localizes into the elementary quanta of light).

To prepare our material for electron diffraction, we first grow a single-crystal (*24*) of stoichiometric composition (see Methods) and then cut out a free-standing membrane with a thickness of ~200 nm using a focused ion beam (Suppl. Fig. S2). The surface normal of the membrane is aligned to the b-axis of the rutile-like crystal in order to make the electron diffraction pattern sensitive to the vanadium dimerization along the c-axis (*9*, *12*). Electron energy loss measurements of the metal phase's bulk plasmon (Suppl. Fig. S3) and measurements of diffraction as a function of temperature (Suppl. Fig. S1) show the phase transition and its sharp hysteresis curve (Fig. 2c). The measured transition temperature varies by less than 0.3 K when we scan a focused electron beam through the laser-excited area (Suppl. Fig. S4). In electron diffraction, we observe only one of the four possible low-temperature twins (*25*).

The light pulses to be absorbed are derived from a femtosecond laser system (Carbide, LightConversion). The pulses have a repetition rate of 300 kHz, a duration of 250 fs, a wavelength of 1030 nm and a photon energy of 1.2 eV, well above the band gap of ~0.6 eV of $VO_2$ (*7*). Adjustable pulse energies in the range of 10-500 pJ are obtained with a pair of polarizers and a parabolic mirror with $f$ = 2.7 mm focuses the laser beam to a spot diameter of 1.8 μm × 1.1 μm on the specimen (Suppl. Fig. S5). For probing the crystal structure as a function of time delay, we apply an ultrafast electron microscope (JEM F200, JEOL) with Schottky field emitter source (*26*) in diffraction mode. We use less than one electron per pulse to avoid space charge effects (*27*) and apply an imaging electron energy filter (CEFID, CEOS) to reject inelastically scattered electrons.

Figure 2d shows the electron diffraction pattern at negative pump-probe time delays, that is, before any absorption of laser light. We see all characteristic Bragg reflections of low-temperature/monoclinic $VO_2$ (*10*, *12*). The strongest Bragg spots are called rutile spots because they appear in both the low-temperature/monoclinic and the high-temperature/rutile-like phase, having a ~30% higher intensity in the high-temperature/rutile-like phase (*12*). Between these peaks, we see the expected additional spots at half distance between the rutile Bragg spots due to vanadium-vanadium dimerization and unit cell doubling in the monoclinic phase (*7*). These spots disappear completely when the material transforms into the high-temperature/rutile-like phase (*9*, *10*, *12*). The clarity of the diffraction pattern shows that our material is single crystalline and in a homogeneous monoclinic phase throughout the entire probed area.

After the laser hits, we record the integrated intensities of the various Bragg spots and the diffuse scattering in the areas between them as a function of the pump-probe delay to search for global or local phase transition effects. We normalize all measured intensities by the total diffracted intensity, excluding the direct beam (see Methods). In a first experiment (Fig. 2e-g), we apply a laser power of 25 µW or 80 pJ per pulse, corresponding to an excitation fluence of ~3.7 mJ/cm$^2$. Inside of the illuminated volume, the resulting energy deposition (Suppl. Fig. S5) is insufficient to reach the transition temperature (see Methods) and even further away from being able to overcome the latent heat (85 meV per rutile unit cell). We therefore do not create a superheated or transformed material. Consequently, for long pump-probe delays, all diffraction intensities remain almost unchanged (Fig. 2e-g). On picosecond time scales, however, we observe substantial variations of the diffraction intensities, while the Bragg spot widths remain unchanged (Suppl. Fig. S7). Figure 2e shows the measured intensity changes in the monoclinic spots. We see a transient decay by ~7% before the intensity returns to its original value at ~5 ps. Figure 2f shows the dynamics of the rutile spots. The intensity quickly increases and then returns to its baseline within a few picoseconds. Figure 2g shows the dynamics of the diffuse scattering, that is, all electrons that are diffracted into the initially empty regions of the diffraction pattern. The intensity of this diffuse scattering quickly rises by ~5% before it returns back to its initial value within ~5 ps. This transient diffuse scattering exceeds by a factor of ~10 the amount that could be expected from thermal phonons in a homogeneously laser-heated material (see Methods). We see no alignment of the diffuse scattering to the crystallographic axes of $VO_2$ (Suppl. Fig. S12).

A diffuse scattering of such magnitude and homogeneity can only originate from a random, non-ordered set of point-like scatterers, that is, an irregular pattern of small transformed spots. A more regular atomic displacement, for example a random delocalization of all atoms by temperature, would produce much less effect and no increasing signal in the rutile Bragg spots (Methods). We conclude that random and tiny parts of our crystal transform for a short time from monoclinic to rutile as small and discrete domains, although the deposited energy density is too low to overcome the latent heat and create any lasting effect. The electromagnetic energy of the laser beam is therefore not absorbed homogeneously within the entire illuminated volume (Fig. 1a). Instead, we create individual domains of excited material in a local way (Fig. 1c). In these distinct spots, the local energy density becomes larger than average and triggers the structural phase transition in selected unit cells or tiny domains. On picosecond time scales, heat dissipation then transforms these local domains back to the original low-temperature phase, in which the macroscopic material must end up after sufficiently long time (Suppl. Fig. S8 and Suppl. Movie 1). Consequently, all three signals return to their original values.

Figure 3 shows a series of additional time-resolved diffraction data for a large range of laser excitation strengths. For low fluences (yellow to light orange), we observe almost no lasting monoclinic or rutile long-time change (>3 ps), although there is some transient diffuse scattering. For intermediate fluences (dark orange to magenta), we also measure almost no long-term change, but the transient state with reduced monoclinic intensity and increased rutile and diffuse intensities lasts for a longer time (several ps). For higher laser fluences above the transition threshold (violet to black), we observe a substantial and lasting transformation of monoclinic into rutile material (*8–12*), but there is in all cases also a transient disorder (Fig. 3c) that emerges within the time resolution of our experiment and lasts for several picoseconds. These results show that the material initially transforms in a localized way, irrespectively of whether the total absorbed laser energy is below, at, or above the energy threshold needed for eventually creating a macroscopically stable state at later times.

In order to clarify whether the observed disorder is a spontaneous break of random, individual V-V bonds or whether the much larger energy of individual photons becomes localized, we estimate from the experiment the absolute number and the size of the transformed spots. For the experiment of Fig. 2, the absolute density of absorbed photons in our material is ~3.5% per rutile unit cell (see Methods) and we transform about ~4.5% of the rutile unit cells (see Fig. 2f). The amount of structural change directly after laser incidence (~0.3 ps) scales linearly with photon density (Suppl. Fig. S9). The measured rise of diffuse scattering and loss of monoclinic intensity is not proportional but follows a path (Suppl. Fig. S6). Because diffuse scattering is roughly proportional to the surface and monoclinic loss to the volume of the transformed spots, we can roughly infer their size (see Methods). Directly after laser excitation (200-300 fs), the estimated volume is ~0.4 nm$^3$ (Suppl. Fig. S6) which is much larger than one unit cell with one V-V dimer (~0.06 nm$^3$). The energy cost to form such a domain is $E_{\text{cluster}} = E_{\text{L}} + E_{\text{heat}} + E_{\text{J}}$, where $E_{\text{L}} \approx 571$ meV is the latent heat of the domain, $E_{\text{heat}} \approx 300$ meV is the energy cost to heat the domain up to the transition temperature, and $E_{\text{J}} \approx 296$ meV is the energy of the domain wall between the two different phases at the cluster surface (*28*). We obtain $E_{\text{cluster}} \approx 1.17$ eV, similar to one photon energy (1.2 eV), ~20 times more than the energy needed to switch one V-V dimer (60 meV), or ~15 times more than the unit cell equivalent of the latent heat (85 meV), or ~50 times more than the energy of the promoting 6-THz phonon at ~25 meV (*15*). We therefore argue that the observed localization and break of translational symmetry is a genuine photonic effect and not a statistical break of random V-V bonds. According to the above estimations, each photon of the laser light produces approximately one localized spot of multiple switched unit cells.

The check whether this picture is compatible with the full ultrafast time traces of our measurements, an information not used so far, we simulate the cooperative flipping of adjacent unit cells between the two stable phases with a kinetic Monte Carlo simulation (*28*) and couple it with a simulation of heat flow (*29*). We model the localized absorption of individual photons by depositing their energy into random unit cells with a probability that is proportional to the intensity of the laser beam and the phase-dependent absorption coefficient of the material. The simulated phase maps as a function of time are then converted into atomic positions (*12*) and the resulting time-dependent diffraction patterns are analyzed in the same way as in the experiment (see Methods). Figure 3d-f show the results for the monoclinic, rutile and diffuse scattering. All relevant features of the experiment are qualitatively and quantitatively reproduced. Much more and smaller excitations or much less and larger ones would produce different dynamics in the time domain. This result confirms that each photon causes its own local phase transformation when absorbed in the initially untransformed material.

Figure 3g depicts the simulated physics in space and time; see also Suppl. Movie 1. At $t = 0$, when the laser hits, we observe the appearance of tiny, localized spots with sufficiently high local temperature to quickly transform several adjacent unit cells from the initial low-temperature/monoclinic into the high-temperature/rutile phase. Consequently, diffraction intensity is transferred from the monoclinic to the rutile Bragg spots (Fig. 3a-b). However, only a small fraction of all unit cells is switched at spot diameters of ~1 nm (Suppl. Fig. S8). These localized phase changes produce homogeneous diffuse scattering with no relation to the crystallographic axes of the material (Fig. 3c). Afterwards, the spots dissolve by heat dissipation and the material starts to convert back to the original low-temperature/monoclinic phase at only slightly elevated temperature. Already at ~3 ps, the material arrives at a mostly homogeneous state with almost equilibrated temperature and no more remaining high-temperature/rutile-like domains, as demanded by thermodynamics and the conservation of total energy. If the excitation would be homogeneous throughout the material, no ultrafast phase change would be possible below the phase transition threshold, in contrast to the experiments (Fig. 2-3).

For higher laser fluences (Suppl. Movie 1), but still at an average energy density below the full transition, we create a denser local composition of tiny domains that slowly converts into a more consolidated geometry with larger domains and fewer domain walls. Therefore, in the experiment, the long-time ratio of monoclinic to rutile-like material remains almost constant (black curve in Fig. 3d-e at >2 ps) although the diffuse scattering continues to decrease (black curve in Fig. 3f at >2 ps). For a short time, where we have localized metallic/rutile spots (*4*, *15*) but no long-range rutile order, $VO_2$ may therefore appear to be monoclinic and metallic (*4*, *10*, *21*, *30*, *31*) without the need to invoke a genuine monoclinic-metallic crystal phase (*10*, *30*, *31*). At even higher fluences, where the entire material is excited above threshold (*9–12*), almost all V-V dimers obtain a photon and the resulting atomic dynamics becomes a ballistic phenomenon (*9*) in which initial noise by temperature creates characteristic fluctuations in the V-V distances (*4*, *11*, *21*).

The phase transformation in $VO_2$ does not only change the atomic geometry of the unit cell, but also involves an insulator-to-metal transition (*6*, *7*). Consequently, we should also observe the signature and dynamics of local phase transformation in an all-optical experiment (Fig. 4a). We use the same weak pump pulses below phase transition threshold as in the electron diffraction experiment but now measure the time-dependent electrical properties of insulating or metallic $VO_2$ with ultrashort mid-infrared laser pulses at a center wavelength of 6.5 µm (*32*), far below the band gap and away from electronic band structure effects (*7*, *8*, *14*). The pump and probe pulses impinge in collinear geometry on a single-crystal of $VO_2$ (*24*) and the reflected mid-infrared intensity is recorded as a function of pump-probe delay (see Methods). Figure 4b shows the results. At low fluences (yellow to orange), we observe a quick increase in reflectivity that rapidly decays back to the initial value on a timescale of ~1 ps. At intermediate fluences (orange to red), the differential reflectivity first increases rapidly, then decays, and ends up at a finite value. At higher fluences above threshold (violet to black), the initial spike slowly disappears, and we only observe a long-lived increase of reflectivity lasting for many picoseconds. A slight and slow increase of reflectivity after several picoseconds is caused by heat diffusion from the surface into the bulk of the material. The amount of initial metallic signal directly after laser incidence (~0.3 ps) scales linearly with photon density (see Suppl. Fig. S9). In the simulations (Fig. 4c), we apply the same model and parameters as in the simulated diffraction results but calculate the change of reflectivity from the two refractive indices (*33*) with an effective medium approach (see Methods). All relevant features of the experiment are reproduced, supporting the results from the diffraction experiment.

In a third experiment (Fig. 4d), we investigate the peak electron temperature of our laser-excited $VO_2$ crystal by recording the power of the emitted thermal radiation (*34*) above and below threshold. We use an almost single-photon-sensitive spectrograph to measure the high-frequency tail of the Maxwell-Boltzmann distribution in the range of 600-700 nm, where we are only sensitive to electronic temperatures above ~1000 K (see Methods). Figure 4e shows the results. In the range of 0-6 mJ/cm$^2$, we see no substantial emission at all, because the photon energy is immediately used up by breaking several V-V dimers, leaving a cold electron gas that does not radiate. A small quadratic signal is caused by light that is re-absorbed by unit cells that have already switched by previous photons in the pulse, a clear signature of the expected effect. Above 10 mJ/cm$^2$, we see a linear rise of the emitted thermal radiation with laser fluence. In this regime, almost the entire material is switched; excess photons cannot anymore trigger the phase transition and instead heat up the electrons in the material which in turn emit visible light. Interestingly, the spectral shape of this thermal radiation (Fig. 4f) does not change with the applied photon density and the amount scales linearly with photon density, although the high-frequency tail of Maxwell-Boltzmann radiation should increase exponentially with temperature (*35*). We argue that more absorbed photons and photon-generated electrons do not directly produce a homogeneously heated material but initially more islands of equally-hot material, causing a linear emission increase. If the laser beam would be absorbed homogeneously as a wave, more laser light would simply create a hotter material, resulting in approximately exponential growth of the emitted high-frequency radiation (*35*) and changes of the spectrum towards shorter wavelengths (*36*), contrary to what is observed (Fig. 4e-f). Again, the simulations with localized photon energies (black line) reproduce all measurements.

The combined results from these three independent experiments and their agreement to one simple theory for all parameters above and below the phase transition threshold show that the absorption of laser light in $VO_2$ creates local hot spots at the expense of no relevant excitation elsewhere. This localization and its structural consequences can occur as fast as within femtoseconds, and the resulting inhomogeneity then persists as long as it takes for heat diffusion to provide a new equilibrium. Absorption of coherent laser light by a homogeneous and crystalline material therefore first breaks and then restores translational symmetry on ultrafast and atomic scales (see Suppl. Movie 1). Even at very low laser intensities, far below the phase transition threshold, there is still a tiny but reproducible amount of switched domains of nanometer size. Furthermore, our results indicate that the number of localization events is directly linked to the number of photons in the absorbed electromagnetic wave, not to the number of energetically available V-V bonds. Each incoming photon therefore creates its own localized domain of switched material.

What physics can localize the incident coherent laser wave in a single-crystalline material into nanometer dimensions? A first mechanism might be crystal defects which could provide molecule-like absorption properties that defy a band-structure description and thus can provide deterministic seeds for the creation of localized domains. However, the observed number of switched domains is substantially higher than the defect density in our material (see Methods) and also scales linearly with laser energy (Extended Data Fig. S9), an effect that is not expected for defect-mediated absorption effects (Methods). In addition, the match of three independent experiments (electron diffraction, optical spectroscopy, thermal radiation) on three independently produced materials to one simple theory makes it unlikely that defects have a major effect. Second, $VO_2$ is a transition metal oxide with localized d-bands, that is, electronic states with charge density strongly concentrated around V atoms. Still, these states form a Bloch-wave band that extends, in principle, over the entire crystal, even in the presence of defects (*37*). It is, however, possible that thermal fluctuations and temperature effects cause a condensation of the excitation into selected V-V bonds (*4*, *15*). In this theory (*4*), we could expect roughly as many broken V-V dimers as energetically available, not fitting well to what is observed. Third, we may argue with decoherence theory (*1–3*): The absorption of light in a strongly correlated material couples the electromagnetic wave very efficiently to multiple kinds of electrons, phonons, plasmons or related quasiparticles. These multi-dimensional couplings may provide enough possibilities and phase space volume to form a pointer state, that is, a real-space observer for an initially delocalized photonic state. The crystal then becomes a wave-function-collapsing measurement device for the photons, like a phosphor screen in an electron microscope is a wave-function-collapsing measurement device for delocalized beam electrons. In this picture, each photon would create exactly one switched spot, as indicated by our experiments and theory. Although defects and temperature might still contribute, and be necessary as fundamental symmetry breaks, it seems likely that laser light can collapse into photons, the elementary quanta of light, when absorbed by a strongly correlated material such as $VO_2$.

Independently of the exact mechanism and number of domains, we see clearly in the data that the absorption of light by a complex material can–for short times–produce local energy densities that are much larger than the average energy density of the incoming light. In the present experiments, the energy density in the randomly switched spots is up to ten times higher than deposited on average. Processes in a material can therefore be initiated that are impossible to achieve if the radiation would be homogeneously absorbed. We believe that this discovery is potentially useful for material processing, ultrafast spectroscopy and optical technology.

**Acknowledgments:** We acknowledge financial support by Evangelisches Studienwerk e.V. and Deutsche Forschungsgemeinschaft via SFB 1432. We thank Matthias Hagner for support with lamella production, Bastian Jehle, Tobias Morgen and Andreas Himmelsbach for assistance with calorimetry, and Bastian Jehle, Marco Genovesi and Anton Glaskow for help with ICP measurements.

**Data availability statement:** All data are available from the corresponding authors upon reasonable request.

**Author contributions:** All authors performed research and wrote the paper.

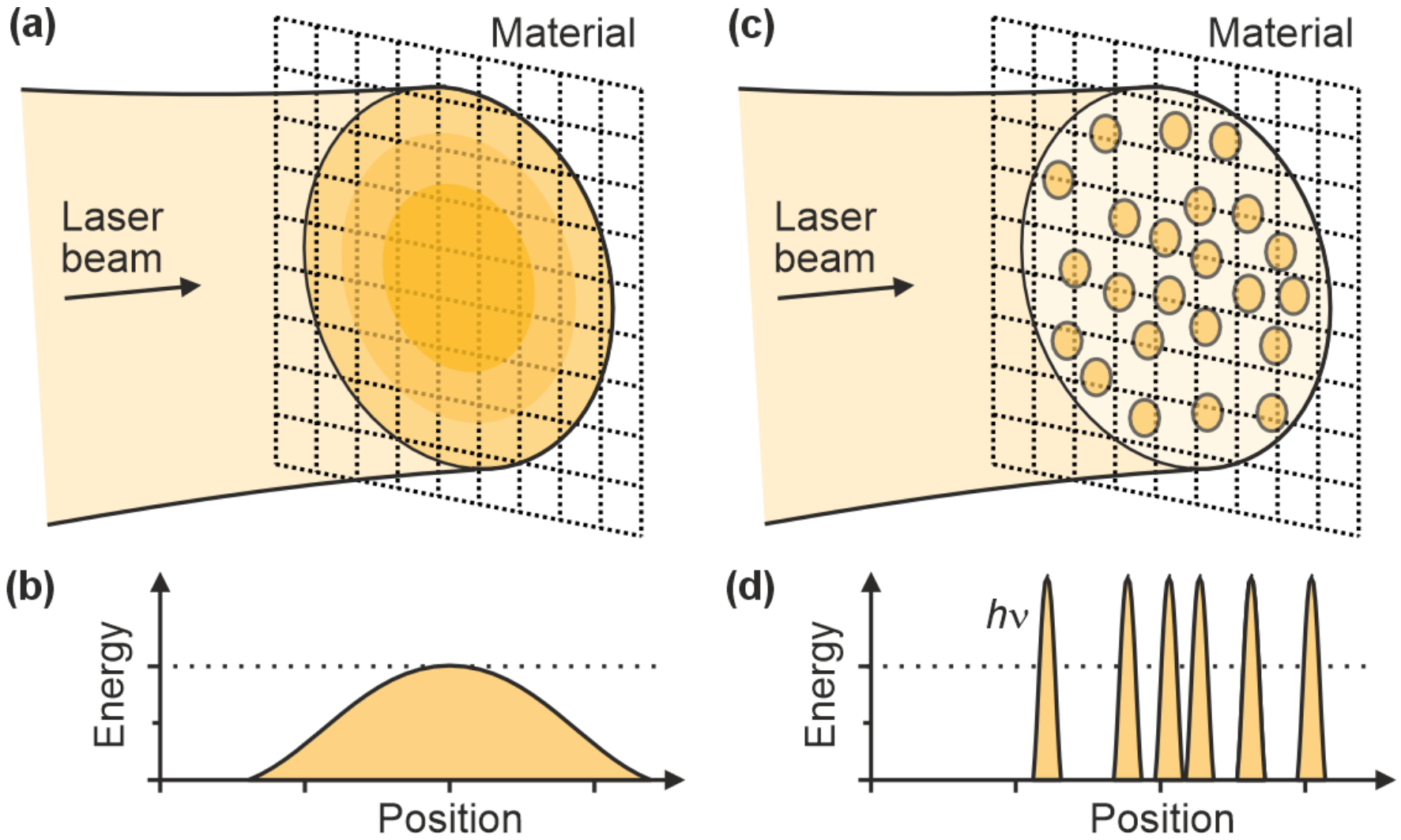

**Fig. 1. Cases for the absorption of laser light by a crystalline material.** (a) A coherent laser beam (orange) hits a crystalline material (dotted lines). Energy is deposited in form of a large, homogeneous spot. (b) The absorbed energy density is proportional to the laser beam profile. (c) Alternatively, the laser beam hits a crystal and deposits its total energy as individual, local excitations (orange spots). (d) The absorbed energy density is a distribution of local spikes with significantly higher peaks than in the wave (dotted line).

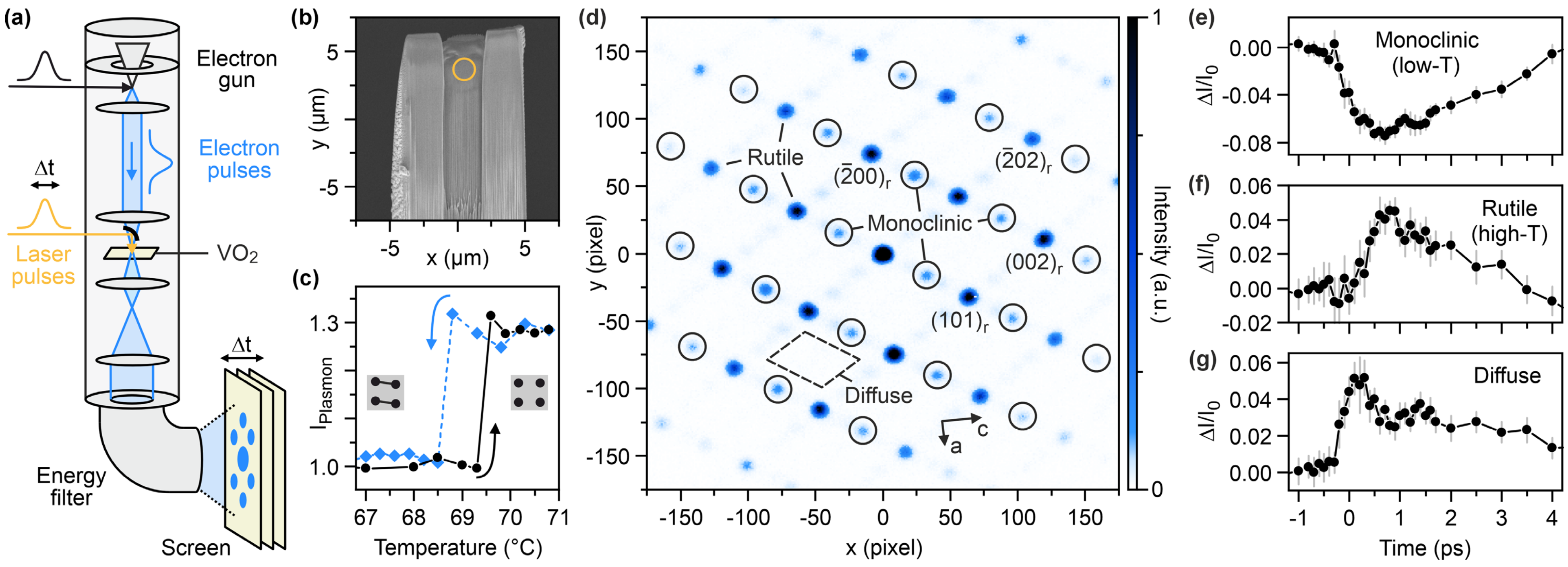

**Fig. 2. Ultrafast electron diffraction of localized absorption effects.** (a) Concept of the experiment. A weak laser beam (orange) hits a single-crystal of vanadium dioxide ($VO_2$) and triggers its first-order phase transition from monoclinic to rutile. Femtosecond electron pulses (blue) probe the transient structural change with pump-probe delay $\Delta t$. (b) Scanning electron microscopy image of the $VO_2$ lamella. (c) Hysteresis curve, observed via the metallic plasmon peak in the high-temperature/rutile phase. Insets: figurative crystal structures with or without vanadium dimers. (d) Electron diffraction pattern of $VO_2$ before laser excitation. We observe all expected Bragg spots. The dashed area depicts regions without Bragg spots that are used for diffuse scattering analysis. Miller indices are given for the rutile phase. (e) Time-resolved intensity $\Delta I/I_0$ of the monoclinic Bragg spots for a laser power that is too weak to reach transition temperature or overcome the latent heat. (f) Time-resolved intensity $\Delta I/I_0$ of the rutile Bragg spots. The slight delay is disucssed in Suppl. Fig. S6. (g) Time-resolved intensity $\Delta I/I_0$ of the diffuse scattering. The magnitudes and dynamics of these results show that the laser energy has localized. Error bars (grey) denote two standard deviations.

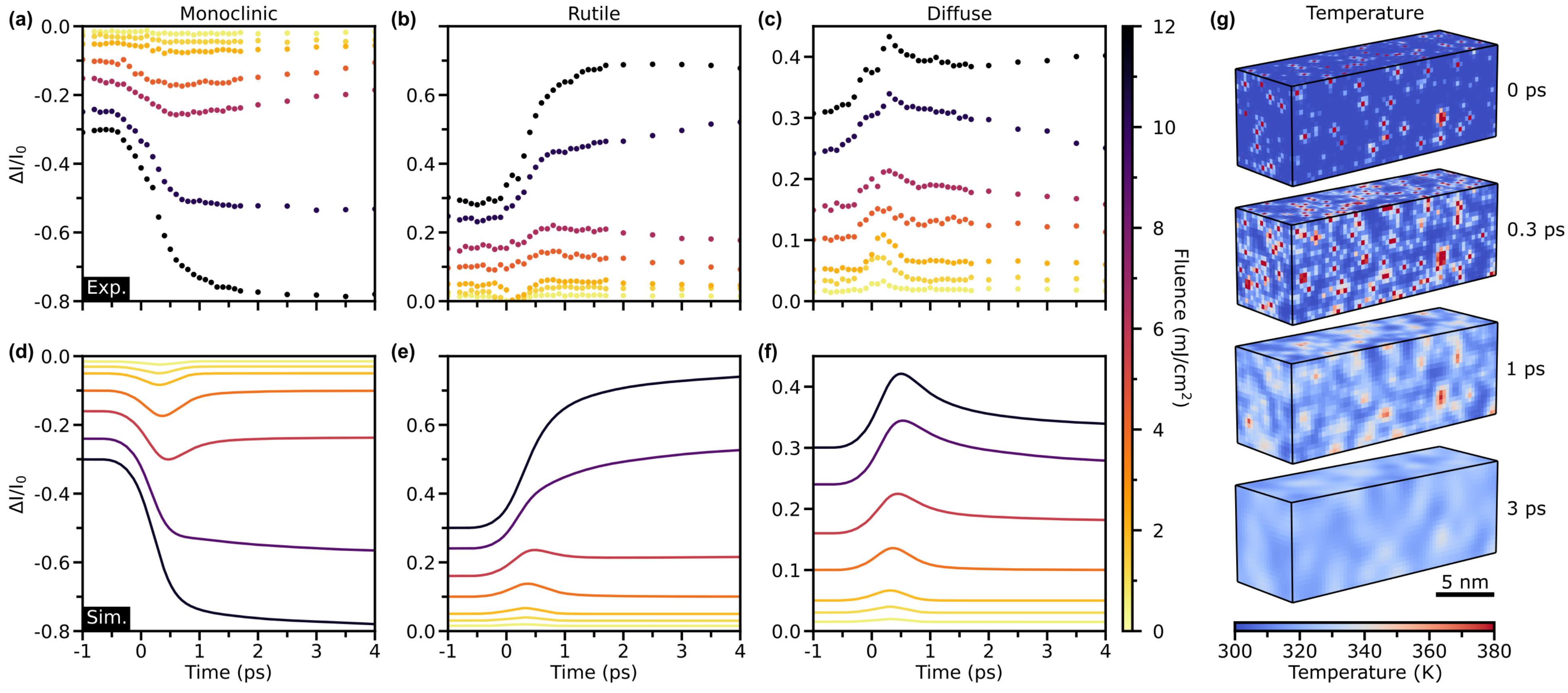

**Fig. 3. Electron diffraction data for extended laser excitation range and simulation results.** (a) Measured time-dependent intensity of the monoclinic Bragg spots. The signal in the short-time drop scales linearly with the absorbed photon density (Suppl. Fig. S9). (b) Measured time-dependent intensity of the rutile spots. (c) Measured time-dependent diffuse scattering. (d) Simulated dynamics of the monoclinic spots. (e) Simulated data for the rutile spots. (f) Simulated diffuse scattering. Data sets are displaced in proportion to applied or simulated excitation strength. The dark orange data is the same as plotted in Fig. 2. (g) Simulated temperature profile within the crystal at selected times. Localized absorption first creates a set of random hot spots at above-threshold temperature (red) that quickly decay back to below-threshold temperature (blue) within few picoseconds. For full data, see Suppl. Movie 1.

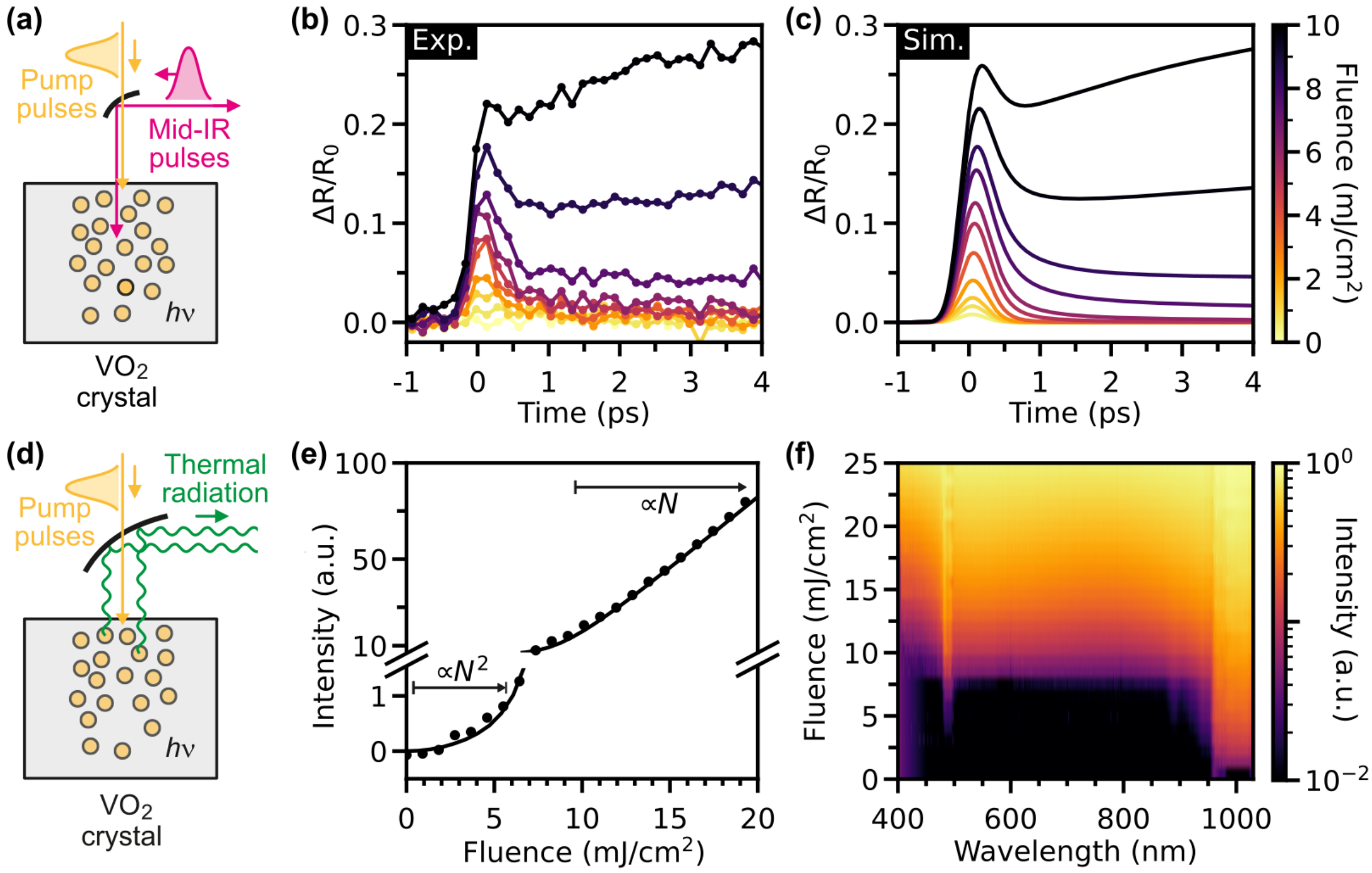

**Fig. 4. Optical mid-infrared and thermal radiation spectroscopy.** (a) Concept of the mid-infrared experiment. A coherent pump laser beam (orange) produces metallic islands (dark orange) that are probed by mid-infrared spectroscopy (violet). (b) Measured optical reflectivity changes as a function of time delay. Colors denote increasing fluences from far below threshold (yellow to orange) over an intermediate regime (orange to red) to an above-threshold regime (violet to black). (c) Simulation results. Localized excitations can, for a short time, switch the electrical conductivity of $VO_2$ even if the applied laser fluence cannot trigger the transition or overcome the latent heat. (d) Concept of the thermal radiation experiment. A coherent pump laser beam (orange) creates localized metallic spots (dark orange) with hot electrons that emit characteristic incoherent light (green). (e) Measured amount of thermal radiation (dots) as function of pump fluence and number of incoming photons $N$. Black line, localization theory. (f) Fluence-dependent spectra of the thermal radiation. Absence of wavelength shifts at growth of total intensity indicates an increasing amount of radiating material at a fixed temperature. The lines around 1000 nm and 500 nm are artifacts from the excitation laser and its second harmonic.

## Materials and Methods

### $VO_2$ Crystals and Sample Preparation

Bulk, stoichiometric single-crystals are grown by thermal decomposition of $V_2O_5$ at 975 °C under an argon atmosphere (*24*). Oxygen and vanadium are to 99.7% single nuclear isotopes. We then use focused ion beam milling (CrossBeam 1540XB, Zeiss) with gallium ions at a current of 20 pA to obtain a freestanding wedge normal to the b-axis of the rutile material. We transfer the wedge onto a support holder (lift-out grid, Pelco) and fix it there with platinum from a gas injection system. Finally, we polish the top of both sides with a current of 50 pA and cut a 2.5×10 µm-wide groove into the specimen with a current of 20 pA, resulting in a thickness of ~200 nm. Supplementary Fig. S10 shows a simulation of this process. The remaining thick material at the sides and the bottom of the groove efficiently removes heat to enable high laser repetition rates (*29*). Scanning electron microscopy images are obtained at an acceleration voltage of 5 keV (CrossBeam 1540XB, Zeiss). Supplementary Fig. S2a shows the sample and Suppl. Fig. S2b its backside. Supplementary Fig. S2c shows a post-measurement cut through the sample at the investigated spot, revealing a thickness of ~200 nm. Black and white in the images denote low and high effective currents of secondary electrons after brightness and contrast optimization.

### Hysteresis measurements

Supplementary Fig. S1 shows measured low-temperature and high-temperature diffraction patterns and a measured hysteresis curve of the structural phase transition in one of our specimens with energy-filtered electron diffraction (JEM-2200FS, JEOL). The electron beam size on the sample is 650 nm, similar to the beam diameter in the ultrafast electron diffraction experiment. A heating-tilting holder (EM-01670SHTH, JEOL) provides temperature steps of 0.1 K. The limited heat conductivity of the sample (*29*) results in a slight offset between the measured transition temperature as compared to the independent plasmon measurements (Fig. 2c). We observe the accompanying insulator-to-metal transition by electron energy loss spectroscopy of a plasmon peak that only exists in the metallic phase (*38*, *39*). Supplementary Fig. S3 shows this peak. We use energy-resolved scanning transmission electron microscopy (STEM-EELS) with a beam diameter of ~10 nm and measure the hysteresis of the plasmon peak on 10×10 positions in a 300 nm × 300 nm sized area. Supplementary Fig. S4 shows twelve of these hysteresis curves. The data in Fig. 2c is the average of the hysteresis curves in the four corners of the measured area. The relative change $I_{\text{plasmon}}$ is the integrated peak intensity divided by the peak intensity at T = 67 °C. In the full measured area, all hysteresis curves are sharp and have the same transition temperture (±0.1 K).

### Ultrafast Electron Diffraction

In our ultrafast transmission electron microscope (*40*), femtosecond electron pulses are generated by photoemission at a Schottky field emitter tip by frequency-doubled laser pulses from a beta barium borate crystal at a wavelength of 515 nm (*41*), focused onto the tip with a $f$ = 300 mm lens. To minimize space charge effects (*42*), we only use ~0.1 electrons per pulse (*27*). The pump pulses, derived from the same laser system, are focused onto the sample under normal incidence with a small, off-axis parabolic mirror that is placed inside of the microscope's twin lens. A small hole in the mirror (diameter 200 µm) transmits the electron beam and only reduces the laser power by an estimated 25%. Diffraction is energy-filtered with a bandwidth of ±2.5 eV and each single electron is detected (*43*) with an event-counting camera (timepix3, Amsterdam Scientific Instruments). The diffraction pattern in Fig. 2d is indexed according to the rutile unit cell; monoclinic Miller indices can be obtained by conversion (*44*). Weak features around the rutile Bragg spots are likely from $Ga_2O_3$ from lamella production, an insulating material with no influence on the measurement. The sample transforms back to ambient condition after each laser shot, confirmed by nanosecond-resolution time traces from residual continuous electron emission (*41*). This ultrafast electron diffraction in a transmission electron microscope marks a substantial advance with respect to previous attempts (*10*, *45–47*), because the pump-probe region is $10^2$-$10^3$ times smaller and the resulting rapid heat removal (*29*, *48*) enables $10^2$-$10^3$ times higher repetition rates.

**Data Analysis and Normalization**

In the ultrafast diffraction experiments, we acquire for each laser pump fluence a series of 7-45 scans of 37 diffraction patterns for different time delays $t$. We randomize the order in which these time delays are measured to cancel potential drifts during each scan. Each image is integrated for 10 s and the total measurement time is 1-5 h per laser fluence. To obtain masks for the various features, we take the mean of all diffraction patterns in the first scan and generate circular masks for the center beam, the rutile and the monoclinic Bragg spots. We select three Bragg spots of each type and fit their positions. We then obtain the mask centers for the other Bragg spots by a regular grid. The position of each circular mask is optimized by maximizing the integrated Bragg intensity. The diameters of the rutile and monoclinic masks are ~2 times larger than the average spot diameters; see Suppl. Fig. S11a. The masks for the diffuse scattering are multiple parallelograms; see Suppl. Fig. S11b. To account for long-term drifts of the electron beam, we track for each time scan the position of the center beam and shift all masks accordingly, while leaving the shape and relative positions of the masks unchanged.

We then sum up for each time delay $\Delta t$ the number of detected electrons (*43*) in the corresponding areas over all scans, leading to measured numbers of electrons $N_x(t)$, where $x = r$ for rutile spots, $x = m$ for monoclinic spots and $x = d$ for diffuse features. To account for changing electron source efficiencies and potential influences of coherent sideband effects in the electron spectrum (*49*), although they are attenuated in normal incidence (*50*), we divide each time trace by the sum of all electrons that hit the rutile, monoclinic or diffuse masks. We obtain $N_x^{\mathrm{norm}}(t) = \frac{N_x(t)}{N_r(t)+N_m(t)+N_d(t)}$. For different laser fluences, we find that the static ratio of the monoclinic to the rutile Bragg spot intensities is not completely the same. We therefore express all data as relative changes $\Delta I_x(t)/I_0 = \frac{N_x^{\text{norm}}(t) - N_x^{\text{norm}}(t<0)}{N_m^{\text{norm}}(t<0)}$. The static monoclinic intensity at negative time delays is the maximum that can migrate into the rutile or diffuse areas. Errors are standard errors of the mean of the normalized intensities at each time delay, averaged over all scans.

The appearance of isolated, localized, tiny spots can best and most unambiguously be proven by scattering events as far away as possible from any crystallographic direction of the raw material. When we alternatively determine the diffuse intensities by using the inverse of the rutile and monoclinic masks, thereby including the $Ga_2O_3$-spots and the V-V dimer line (*11*, *51*), we see the same, only noisier results (Suppl. Fig. S12).

**Ultrafast Mid-Infrared Experiments**

We excite the surface of our single-crystal (*24*) with similar pump pulses as in the diffraction experiments, only from an earlier version of our laser system (Pharos, Light Conversion). The center wavelength is 1028 nm, the pulse duration is 300 fs and the repetition rate is 25 kHz, slow enough to let the sample relax back to ambient temperature between each laser pulse (*29*). To produce mid-infrared probe pulses at a photon energy below any band structure effects (*52*), we generate a supercontinuum in a 10-mm-long yttrium aluminum garnet crystal and amplify the spectral region around 1200 nm to a pulse energy of ~1.4 µJ in a 3-mm-long type-I beta barium borate crystal pumped by the second harmonic of our driving laser. Finally, we use difference frequency generation in a 1-mm-long type-II lanthanum gallium silicate crystal to produce mid-infrared pulses with a center wavelength of ~6.5 µm, a pulse duration of ~80 fs, and a pulse energy of ~50 nJ. A similar setup is described in more detail in Ref. (*32*).

A parabolic mirror with $f$ = 150 mm focuses the mid-infrared probe pulses onto an optically flat crystal surface (*24*) with a spot diameter of 125 µm under normal incidence. A lens focuses the pump pulses in a collinear geometry through a small hole in the parabolic mirror down to a spot diameter of ~450 µm. The reflected probe light is then collected by the same parabolic mirror and guided to the detector by a beam splitter. We employ a balanced detection scheme with two InAsSb photodiodes (P12691-201G, Hamamatsu) to measure the change in reflectivity upon excitation relative to an orthogonally polarized reference pulse that hits the sample before the pump pulse.

**Thermal Radiation Experiment**

We use the same laser system and similar single-crystals as in the ultrafast mid-infrared experiment and observe potential radiation from hot electrons in $VO_2$. We collect the emitted radiation from a 100-µm excited spot with a parabolic aluminum mirror (NA = 0.45) and measure the spectrum as a function of excitation fluence a with a Czerny-Turner-style monochromator (SpectraPro 300i, Acton) together with an electron-multiplying CCD camera (iXonEM+ 897, Andor) with a quantum efficiency of ~90% in the visible range. Supplementary Fig. S13 shows the fluence dependent raw spectra as recorded with the spectrometer. The spectra in Fig. 4f are normalized to the spectrum at maximum fluence (25 mJ/cm$^2$). In Fig. 4e, we plot the integrated signal between 600-700 nm. To account for long-term drifts, we measure each fluence three times in randomized order at an integration time of 300 s. We observe the same scaling below and above 515 nm, excluding a potential contribution from photoluminescence (*53*) from residual second harmonic laser light.

**Quantitative Cooperativity Simulations**

Both simulated datasets (diffraction and reflectivity) are obtained with a parameter-free kinetic Monte Carlo simulation, coupled to a heat diffusion model which is solved using the finite difference method. The kinetic Monte Carlo simulation identifies the two phases of our material with the two spin states $\sigma_i = \pm 1$ of the Ising model (*54*). We assume a simplified, cubic unit cell with a dimension of 0.5 × 0.5 × 0.5 nm$^3$, similar to the dimensions of the monoclinic unit cell with two V-V dimers. We obtain the Hamiltonian (*28*) $H = -\frac{1}{2}\sum_{i,j\in NN} J \cdot \sigma_i\, \sigma_j + \sum_i h_i(T_i)\sigma_i$, where $J$ = 13.8 meV is the cooperative interaction strength between neighboring $VO_2$ unit cells (*28*) and $h(T) = L\frac{T_t - T}{T_t}$ is an external temperature field. The latent energy per simulated unit cell with two dimers is $L$ = 180 meV, obtained from the latent heat of 50 kJ/kg (*6*, *55*). Slight mismatch to the real latent heat of 2 × 85 meV per rutile unit cell originates from the approximated theoretical, cubic unit cell size of (0.5 nm)$^3$ = 125 Å$^3$ versus the real 2 × 61.85 Å$^3$ in monoclinic $VO_2$. Each calculated phase change of the kinetic Monte Carlo simulation absorbs or releases this latent heat. The resulting temperature map $T_i$ then propagates according to classical heat dissipation with the reported heat capacity, thermal conductivity and latent heat of $VO_2$ (*55*, *56*). To simulate the localized photon absorption, we deposit the energy of individual photons as local temperature increase in random unit cells, with a probability distribution following the local laser intensity and phase-dependent absorption coefficient of the material. We thus include the effects of the optical penetration depth (Suppl. Fig. S5c). For diffraction (Fig. 3), we simulate a slab with a front face of 100 × 100 unit cells (50 × 50 nm$^2$) and 400 unit cells in depth (the full sample thickness). For each layer of unit cells, we treat the vanadium atoms as Gaussian spots with a delocalization of $\sigma$ = 37 pm and place them at phase-dependent positions (*12*) on a 5 pm grid. We average the Fourier transforms of all layers to obtain the diffraction pattern. Time-resolved data is obtained with the same masks as in the experiment (Suppl. Fig. S11). Such a kinematic approach is appropriate because we only consider changes of measured intensities and not absolute values. In the reflectivity calculations, we simulate a slab with a front face of 50 × 50 nm$^2$ and a depth of 400 nm, more than twice the pump laser's penetration depth. Optical reflectivity is obtained from the simulated domain structures with an effective medium approach (*57*) and reported values for the refractive index at 6.5 µm (*33*), ~27% in the low-temperature phase and ~72% in the high-temperature phase. In Fig. 4e, we plot the simulated number of photons that are absorbed in already metallic unit cells, because only those can emit substantial thermal light.

**Absence of Crystal Defects**

Crystal growth by liquid diffusion produces stoichiometric crystals with the right amount of oxygen removed; see Fig. 4f of Ref. (*24*). Differential scanning calorimetry (Phoenix 204F1, Netzsch) of a single crystal (Suppl. Fig. S4b-c) shows a one-step transition of the entire macroscopic crystal (volume ~1 $mm^3$, see inset) within 2 K and a latent heat of ~55 J/g, consistent with literature (*6*, *55*). We verify the absence of doping with energy-dispersive X-ray spectroscopy (EDX) in a scanning electron microscope (Gemini 500, Zeiss; with Oxford Instruments EDX detector); see Suppl. Fig. S14. Besides vanadium and oxygen, we only see Al, Si and Ge at levels below 0.1%; other chemical elements are not detectable. We exclude the presence of oxygen vacancies with inductively coupled plasma optical emission spectroscopy (ICP-OES5800, Agilent Technologies). We grind our vanadium dioxide single-crystals into a fine powder, dissolve ~20 mg in 69% nitric acid and dilute the sample with ultrapure water (Milli-Q, Merck Millipore) to achieve a final nitric acid concentration of 3% with an expected vanadium concentration of ~40 ppm. We use an industrial calibration standard (vanadium ICP standard solution, 1000 mg/L, Carl Roth GmbH) in 3% nitric acid. The peak intensity of the optical emission line of vanadium at 309 nm reveals a vanadium concentration of 0.3355±0.0023. The agreement to 1/3 of perfectly stoichiometric $VO_2$ shows that our material has less than 1% of oxygen vacancies.

Doped vanadium dioxide would also have a strongly shifted hysteresis curve that we do not observe (Suppl. Figs. S1 and S4). For example, the transition temperature would shift by >20°K for 1% of tungsten doping (*58*, *59*) and doping with Li, Na, K would change the transition temperature by >40 K at ~1% (*60*); see also Table 1 of Ref. (*61*). Oxygen vacancies (of not-reported amount) would shift the transition temperature by ~20 K (*62*) or completely suppress the phase transition (*63*). Contact with an oxygen-removing ionic liquid under applied voltage can fully destroy the phase transition (*64*). In stoichiometric $VO_2$, oxygen vacancies only appear at temperatures above 375 °C (*65*) but in our experiments we never laser-heat the crystal to above 100 °C. Magneli phases, which we would observe in the diffraction experiments, are reported already for defect densities as low as 0.4 percent (*63*).

Supplementary Fig. S1 shows a measured diffraction pattern with a high-coherence, non-femtosecond electron beam. We see Bragg spot widths of less than 0.095 $nm^{-1}$, limited by the electron beam, showing that the material is homogeneous and has no domains or varying strain. Hysteresis curves scanned over position (Suppl. Fig. S4) show all the same shape. There are no local grains that transform easier or less easy into the high-temperature phase, like in a polycrystalline or nanostructured material. Also, if such domains would exist, they would not cool back to ambient temperature in few-picosecond times (Fig. 2e-g).

The measured linear relation between structural or optical change and the number of absorbed photons (Suppl. Fig. S9) can only emerge if there are very few defects (much less than photons), or very many defects (much more than photons). An intermediate amount would produce different responses for laser pulses with less photons per defect than with more photons per defect, and the measured scaling would become nonlinear.

**Gallium Doping during Lamella Production**

In membrane production, the focused ion beam polishes the $VO_2$ crystal in grazing incidence at 89°. We simulate the Gallium doping during the lamella milling process with the Monto-Carlo program TRIM, a package of SRIM (*66*). Its binary collision algorithm provides a precise and reliable estimation of the implanted ion distribution (*66*, *67*). Typical penetration depths are in the nanometer range and decrease with increasing incidence angle (*68*, *69*). We consider only the final polishing step, because it removes the surface of the lamella and thus erases any potential doping from previous milling steps. For the applied grazing incidence (89°) and an acceleration voltage of 30 kV, Supplementary Fig. S10a shows the resulting ion distribution and Suppl. Fig. S10b the corresponding depth-dependent doping profile. We see that 95% of the ions are absorbed within the first 12 nm and 99% within the first 15 nm. The absolute dose $D$ of incident $Ga^+$ ions per area is calculated from the ion beam current (20 nA), the illumination time for polishing the final 15 nm (17 s) and by the illuminated area $A$ (2.5 × 7.5 μm$^2$) on the specimen according to $D = It/(eA) \approx 115$ atoms/nm$^2$. For any material below 20 nm, we get an absolute effective gallium doping smaller than $7 \times 10^{-3}$ / nm$^3$ ≈ $4 \times 10^{-4}$ Ga atoms per rutile unit cell.

**Exclusion of Debye-Waller effects and surface phenomena**

In Fig. 2, the deposited laser energy only heats up the material by ~20 K. The resulting Debye-Waller effect is less than 0.5 % (*12*) and would affect all Bragg spots in a similar way, diminishing them, contrary to what is observed (Fig. 3b). Therefore, the absorbed laser energy cannot be in random phonons; the corresponding almost-diffusive scattering at ~20 K above room temperature would be too small. Also, such phonons would not cool back in few-picosecond times. A transformation of only the surfaces would not produce substantial diffuse scattering and would also not cool back in few-picosecond times (*29*), contrary to Fig. 2e-g.

**Fluence and Reflectivity**

All reported fluences in this work are incoming power densities of the pump laser in the center of the beam according to $F = \frac{2}{\pi\omega_x\omega_y} \cdot \frac{P_{\text{avg}}}{f_{\text{rep}}}$, where $\omega_x$ and $\omega_y$ are the beam waists of an elliptical focus, $P_{\text{avg}}$ is the average laser power and $f_{\text{rep}}$ is the repetition rate. All reported focus diameters are full widths at half maximum from which $\omega_x$ and $\omega_y$ are obtained by division by $\sqrt{2\ln 2}$. Our lamella sample has a reflectance of 19% and a transmittance of 20% at 1030 nm, consistently obtained by finite time-domain difference simulations and coherent summation of all reflected fields. The reported fluence values have an estimated systematic error of ±20%.

**Absolute Photon Density and Cluster Size**

We report a quantitative assessment of the measured data in Fig. 2e. At an incoming fluence of $F \approx 3.7$ mJ/cm$^2$ and a penetration depth of $d$ = 180 nm (*33*), one photon is deposited in the front part of the sample for every $N \approx (1\text{-}R)FV/(dE_{\text{ph}}) = 20$ unit cells, where $V = 61.8$ Å$^3$ is the volume of the rutile unit cell and $R \approx 26\%$ is the surface reflectance from Fresnel reflection (*33*). We may therefore expect a fraction of $1/N \approx 5$ % of the original vanadium dimers to symmetrize. Deeper inside the material this number reduces to ~2% (compare Suppl. Fig. S5c). The total density of photons is ~3.5% per rutile unit cell.

We can estimate the number and size of the rutile domains in our experiment from the ratio of the observed changes in diffuse and monoclinic intensity (Suppl. Fig. S6). In the simulations, we randomly create spherical domains of varying size and density. We move all atoms which are inside these spheres from the monoclinic to the rutile positions (*12*), using a delocalization of $\sigma = 10$ pm (*11*). We analyze the resulting diffraction pattern with the same masks as in the experiment. For better averaging, each simulation is repeated 100 times with new random positions of the spheres. Supplementary Fig. S6 shows the results. For each simulated domain volume, we observe a characteristic ratio of diffuse and monoclinic intensity (dotted lines). The black dots show the measured data of Fig. 2e-f. Directly after laser excitation (200-300 fs) we obtain a domain volume of ~0.4 nm$^3$, corresponding to a radius of $r \approx 0.46$ nm. The domain surface is $A_{\text{cluster}} = 4\pi r^2 \approx 2.63$ nm$^2$. The energy cost to form such a domain is $E_{\text{cluster}} = E_L + E_{\text{heat}} + E_J$, where $E_L = V_{\text{cluster}} \cdot \rho \cdot L \approx 571$ meV is the latent heat of the domain, $E_{\text{heat}} = V_{\text{cluster}} \cdot \rho \cdot C \cdot (70 - 20)°\text{C} \approx 300$ meV is the energy cost to heat the domain up to the transition temperature (see Fig. 2c), and $E_J = A_{\text{cluster}} \cdot 2 \cdot J \approx 296$ meV with $J = 56$ meV/nm$^2$ (*28*) is the cooperativity between the different phases at the cluster surface. We obtain $E_{\text{cluster}} \approx 1.17$ eV. These estimations are accurate to about a factor of two. One photon therefore creates not much less than 0.5 and not much more than 2 spots in the material.

**Momentum**

According to the uncertainty principle, a localized domain at a size of $\sigma_x \approx 0.2$ nm needs a momentum spread of roughly $\sigma_p \geq \hbar/2\sigma_x \approx 500$ eV/c. The associated kinetic energy of an oxygen or a vanadium atom is merely $E \approx \sigma_p^2/2m \approx 10^{-5}$ eV, easily available in the crystal lattice. Local V-V dimerization can occur as soon as single atoms are slightly, classically displaced (*4*), breaking the translational symmetry of the phonon bath. Electrons can also take up this momentum at a kinetic energy of ~200 meV for a free electron. In $VO_2$, the V-V bond is mediated by a d-band (*5*) which is rather flat and provides the necessary momentum at much lower energy.

**Superheating and classical nucleation theory**

Assuming a classical, delocalized energy deposition (Fig. 1a), we would heat up the front surface at the center of the beam by $\Delta T \approx (1 - R)\frac{F_{\text{peak}}}{dC\rho}$, where $d$ is the penetration depth (see Suppl. Fig. S5c), $C \approx 656$ J/kg/K is the heat capacity of insulating $VO_2$ and $\rho \approx 4.57$ g/cm$^3$ (*55*). For the fluence in Fig. 2, we obtain a temperature increase of ~51 K. The front part of the specimen in the center of the beam therefore approaches the transition temperature; other parts see less energy density (Suppl. Fig. S5c). Results at lower fluences (Fig. 3 and Suppl. Fig. S9) are measured far below the classical transition temperature.

Classical nucleation theory describes the spontaneous collapse of an energetically instable domain, for example a superheated material, into a stable phase mixture. However, in the experiment (Fig. 2), we see an ultrafast decay of the localized metallic domains within less than 4 ps (Fig. 2e-g). A superheated material would after nucleation remain in a stable phase mixture for infinite time, at least nanoseconds (*29*). The measured ultrafast creation of V-V dimers for fluences far below the regime of superheating (Fig. 3 and Suppl. Fig. S9) is only possible if the photons create, for a limited time, tiny randomly switched domains, in a mechanism that is fundamentally different from classical nucleation theory.

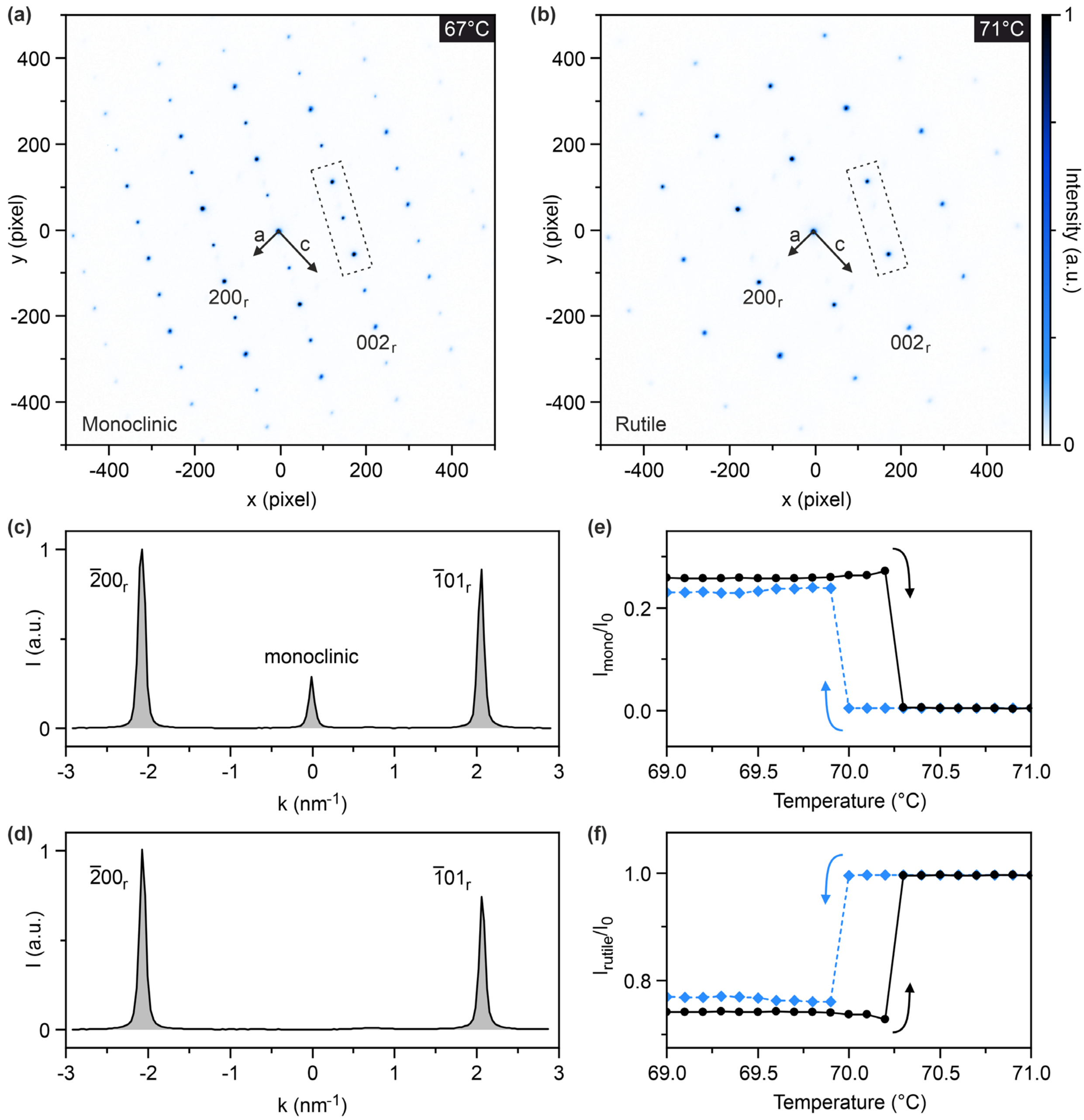

**Fig. S1. Static electron diffraction.** (a) Energy-filtered electron diffraction of our $VO_2$ lamella at low-temperature (67 °C). (b) Electorn diffraction at high temperature (71 °C). The monoclinic spots completely vanish. (c) Line cuts through the dashed regions at low temperature. (d) Line cuts at high temperature. The Bragg spots all have a similar width of $\sim$0.095 nm$^{-1}$, limited by the divergence of the electron beam. (e) Measured hysteresis curve of the monoclinic Bragg spots, using the same mask as in the ultrafast diffraction experiment. (f) Measured hysteresis curve of the rutile spots.

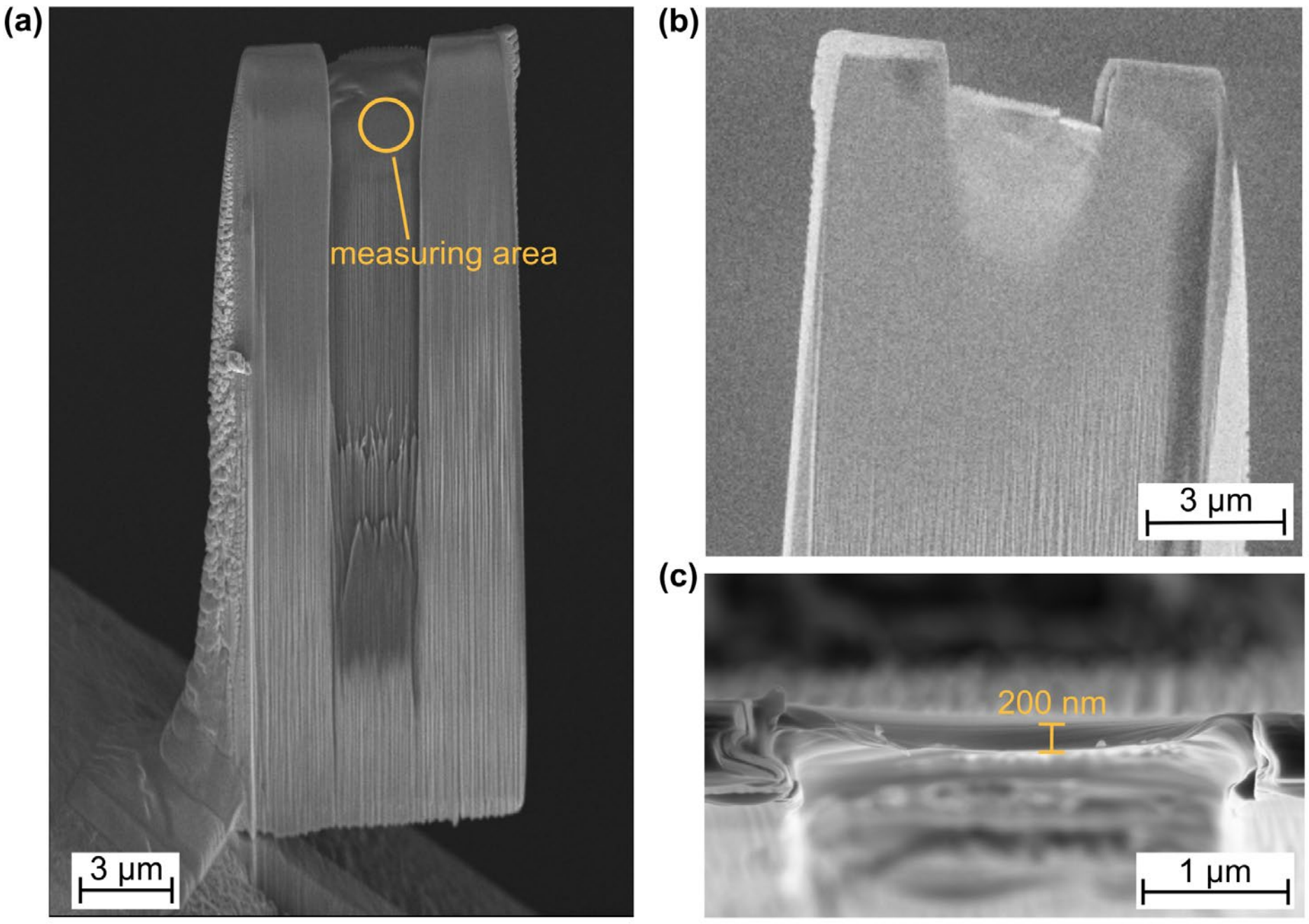

**Fig. S2. Thickness measurement on our lamella sample via scanning electron microscopy.** (a) View on the sample. (b) View on the backside. (c) Top view of a cut through the measurement area. We find a sample thickness of ~200 nm.

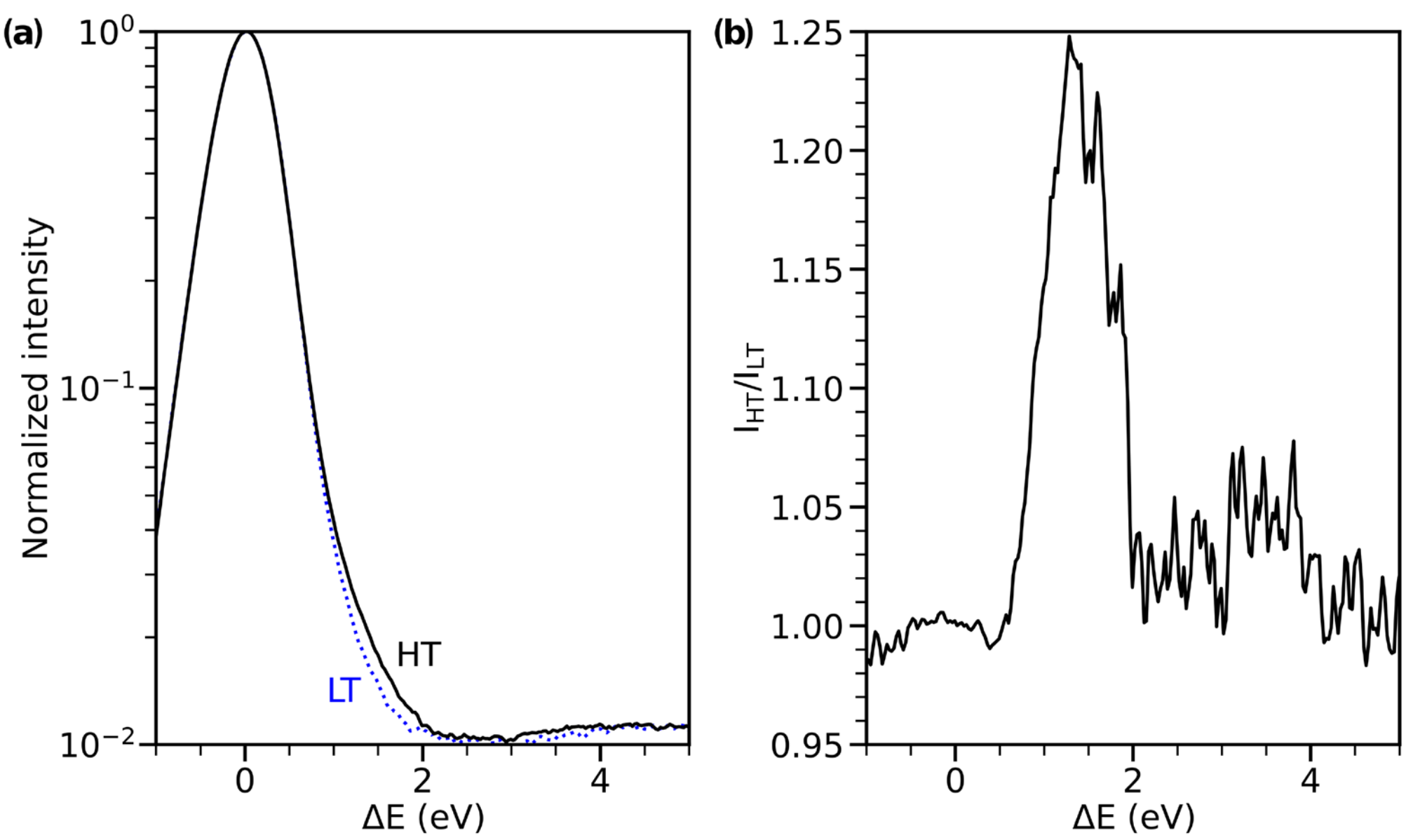

**Fig. S3. Electron energy loss spectroscopy of vanadium dioxide.** (a) Measured electron energy loss spectra of the low-temperature/monoclinic/insulating crystal phase (blue, dotted) and the high-temperature/rutile/metallic phase (black, solid) on one of our $VO_2$ lamellae. (b) Ratio between the normalized intensities $I_{HT}$ in the high-temperature phase and $I_{LT}$ in the low-temperature phase. The peak at ~1.2 eV corresponds to the plasmon loss of metallic $VO_2$ and therefore only exists in the high-temperature phase.

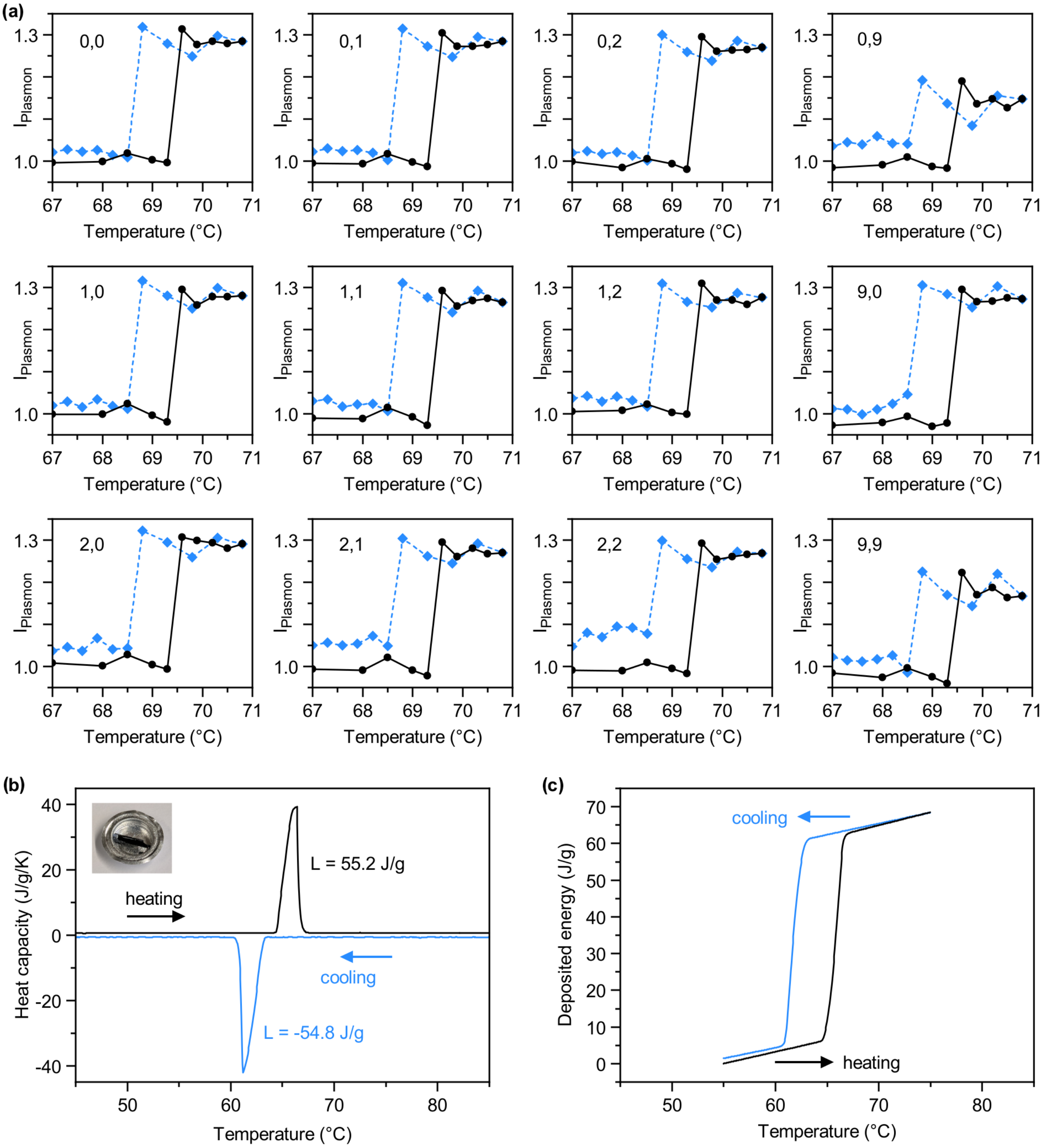

**Fig. S4. Homogeneity of the hysteresis curves.** (a) Using STEM-EELS, we obtain 10 × 10 different hysteresis curves on a 300 nm × 300 nm sized grid using the plasmon peak depicted in Suppl. Fig. S3. The indices in the top left corners denote the position on this grid. All curves show a sharp hysteresis with the same transition temperature. (b) Differential scanning calorimetry (DSC) measurements on our single-crystalline $VO_2$ (volume ~1 $mm^3$, see inset, crucible diameter ≈ 8 mm) reveals during heating (black) and cooling (blue) a full transition of the macroscopic crystal. The baseline is calibrated using the literature value of $C \approx 656$ J/kg/K of insulating $VO_2$ (*55*). (c) Hysteresis curve calculated from the DSC measurement.

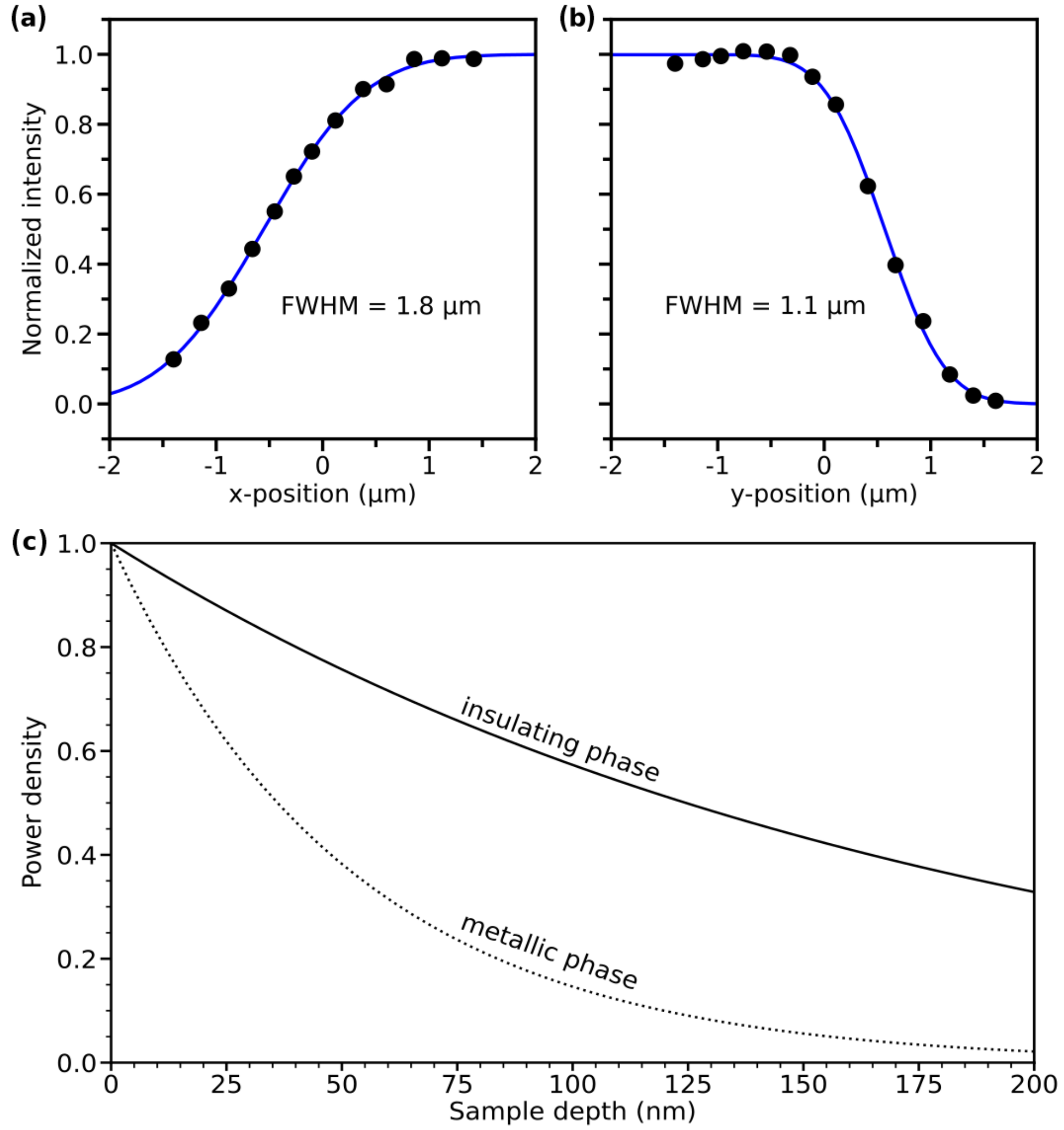

**Fig. S5. Size of the optical pump beam and optical penetration depths.** (a) Knife-edge scans in the x-direction. (b) Knife-edge scan in the y-direction. We scan an edge of the sample through the pump beam and measure the backscattered pump light with a camera that images the sample through the parabolic mirror. The fit with an error function gives us a spot size of 1.8 × 1.1 µm (FWHM, full width at half maximum). (c) Simulated power density profile of our excitation laser beam as a function of depth in the material for insulating/monoclinic (solid line) and metallic/rutile (dashed line) $VO_2$, calculated from the complex index of refraction (*33*). All fluence values and photon densities used in this report refer to the maximum absorbed intensity of the insulating phase at the center of the beam and at the front surface (sample depth of zero). All other parts of the material can only see lower energy and photon densities.

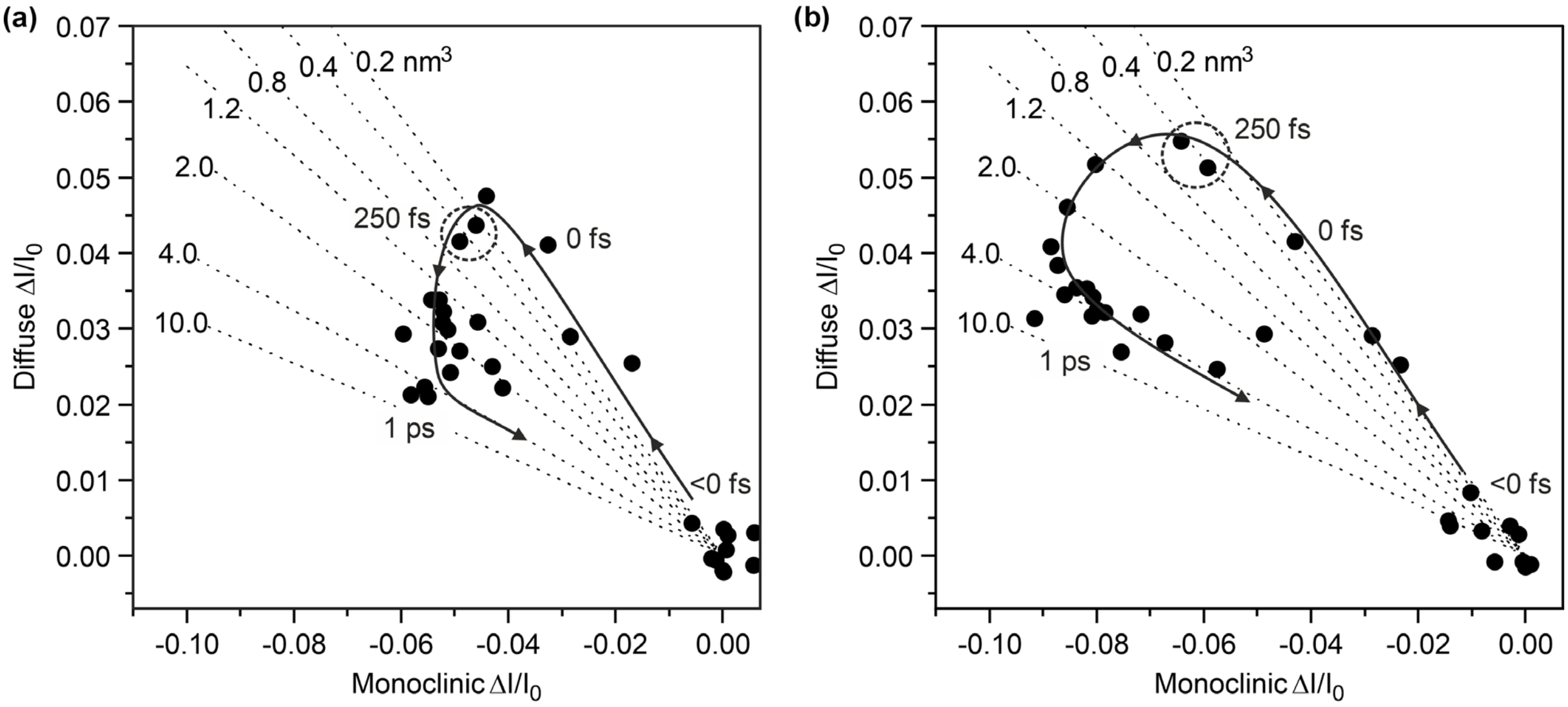

**Fig. S6. Signal correlations and estimation of cluster size.** (a) Correlation plot of the monoclinic and diffuse signals as measured in Fig. 2e-g (dots). (b) Correlation plot of the monoclinic and diffuse signal for a slightly higher fluence of 5.5 mJ/cm$^2$ (dots). The solid arrows show the ordering in time. Dashed lines are the simulated cluster volumes (see Methods). The circles mark the data measured just after full absorption of the laser pulse (200-300 fs).

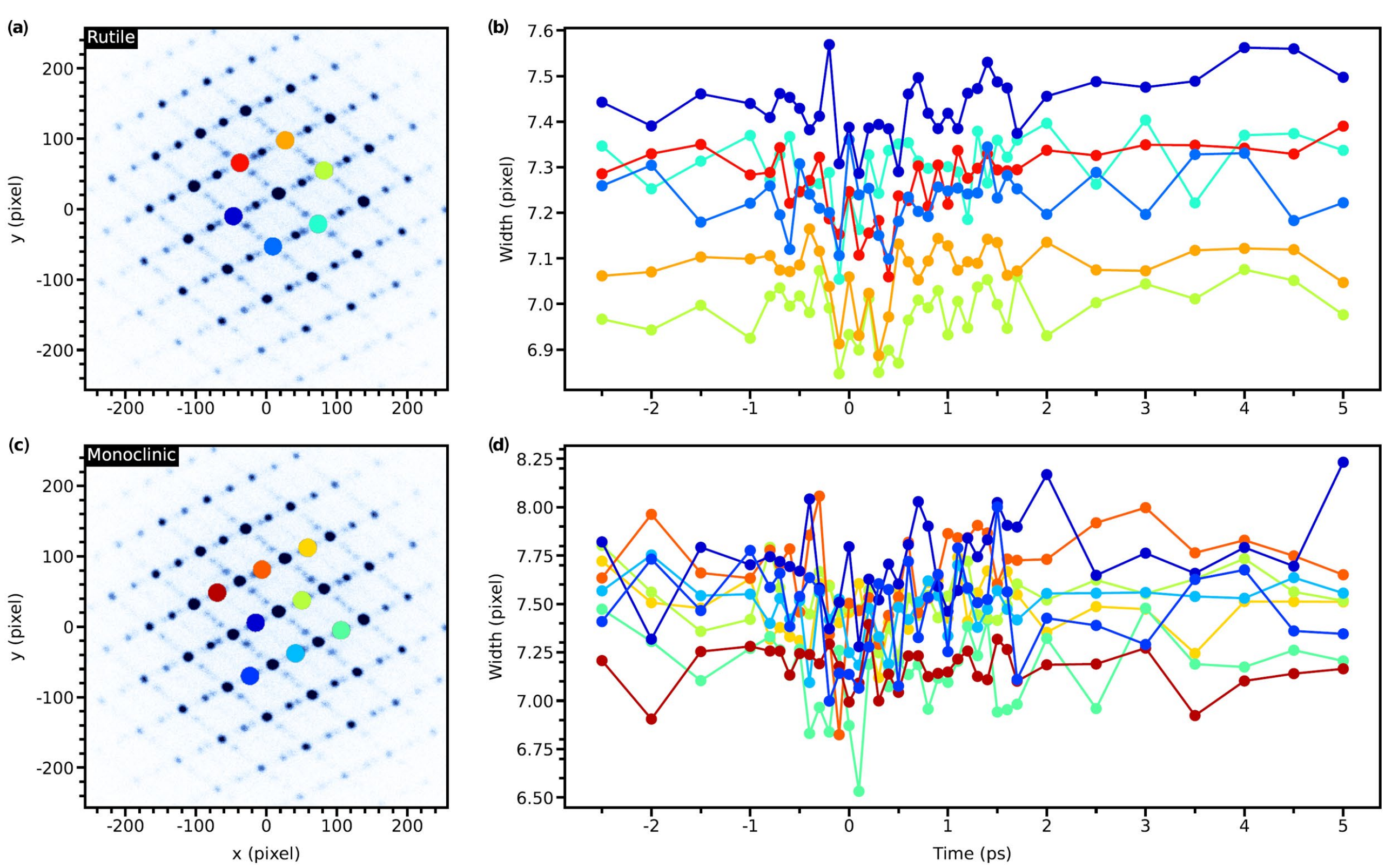

**Fig. S7. Time-dependent Bragg spot widths.** (a) Diffraction pattern and masks for selected rutile spots. (b) Measured Bragg spot widths from Gaussian fits as a function of time. Color of the data refers to the color of the mask. (c) Diffraction pattern and masks for selected monoclinic spots. (d) Measured Bragg spot widths. All rutile and monoclinic spot widths remain constant to about ±2% for all measured times.

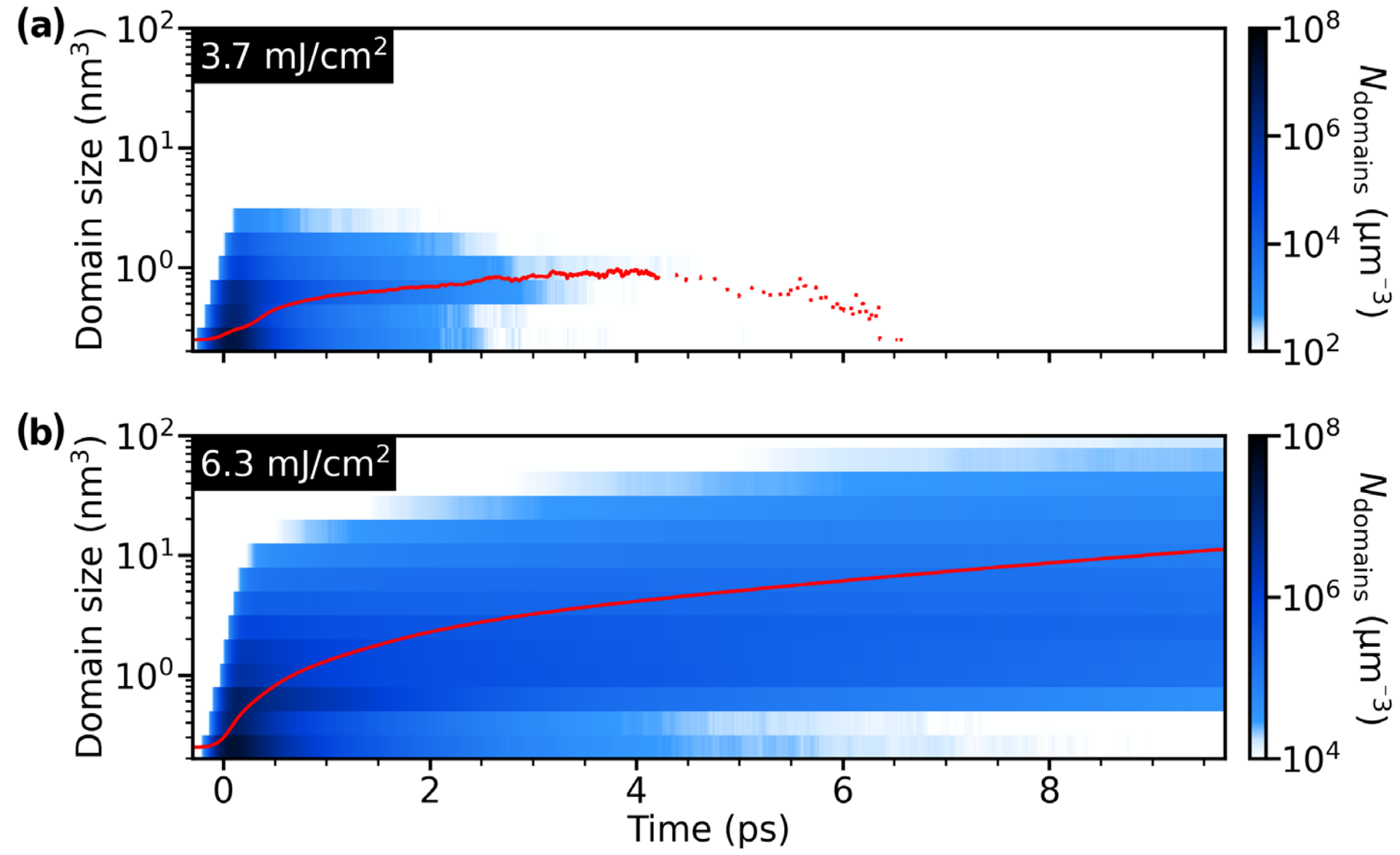

**Fig. S8. Size and growth of the initially localized metallic/rutile domains with time.** (a) Simulated domain sizes and density at an excitation below threshold fluence (3.7 mJ/cm$^2$). During and shortly after laser excitation (0-0.5 ps) we see many small domains (~1 nm$^3$) that appear in time in proportion to the deposited photon density. Afterwards, the typical size increases to ~1 nm$^3$ but at the same time the domains disappear (note the logarithmic density scale) within ~5 ps. The red line shows the average domain size. These results explain the slight delay of the rutile increase in Fig. 2f with respect to the monoclinic decrease (Fig. 2e) and rise of diffuse intensity (Fig. 2g): Diffuse scattering is instantaneous, but rutile Bragg diffraction can only increase after several adjacent unit cells have flipped. (b) Simulated domain sizes and density at an excitation above threshold (6.3 mJ/cm$^2$). We see the initial generation of localized and small domains in a very similar way as in the below-threshold data, but the domains are now dense enough to merge and grow into larger domains that eventually consolidate into a quasi-static phase mixture.

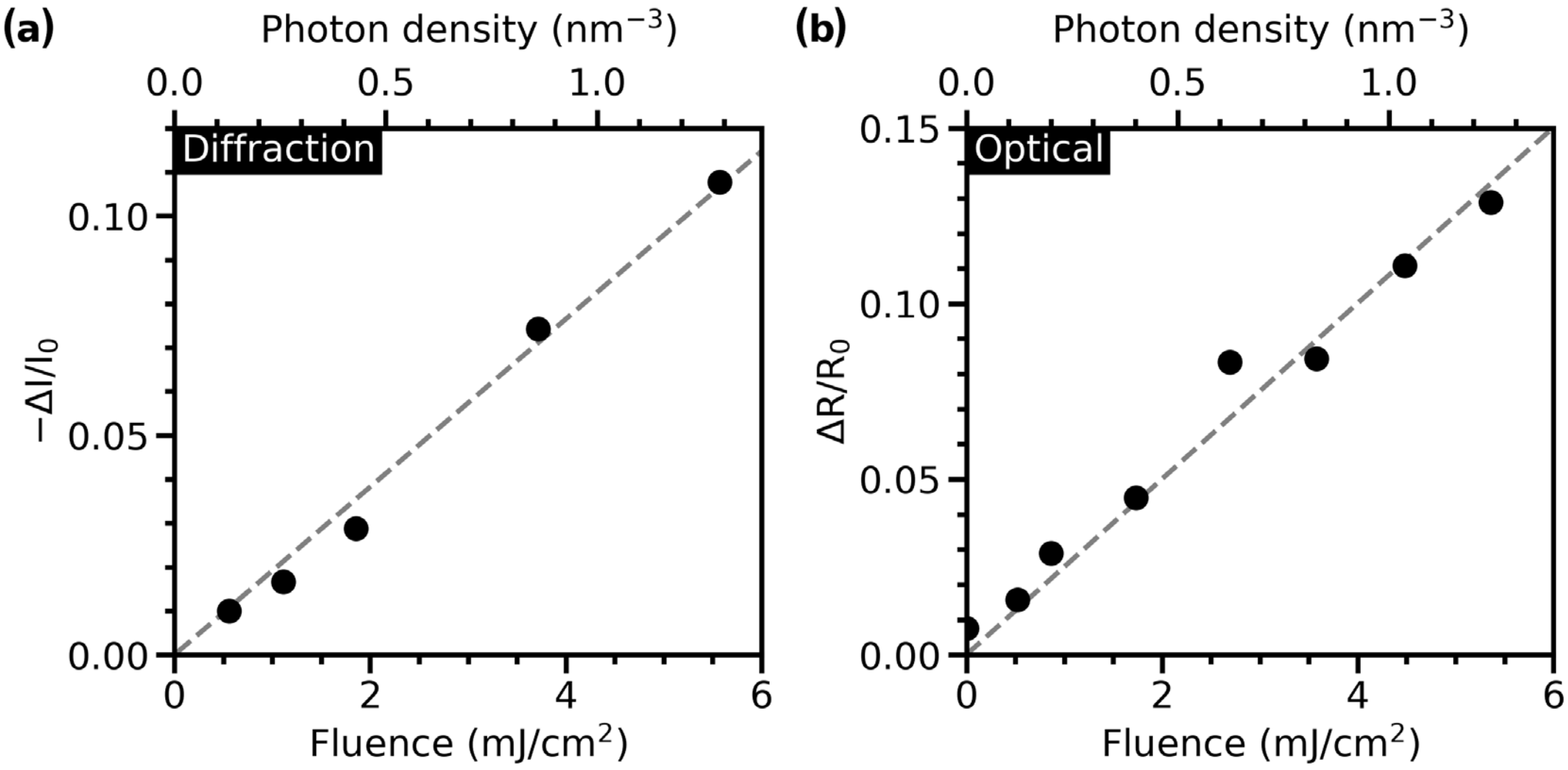

**Fig. S9. Fluence dependence of the ultrafast response.** (a) Measured monoclinic electron diffraction loss (Fig. 3c) at ~300 fs as a function of the applied laser fluence. The photon density on the upper scale is the peak photon density at the front of the sample in the center of the beam. The dashed line shows a linear fit. (b) Measured optical reflectivity changes (Fig. 4b) at ~300 fs as a function of laser fluence. The dashed line shows a linear fit. Both measurements show that the amount of structural disorder (Fig. 1b) or effective metallicity (Fig. 4a) at short times is directly proportional to the number of absorbed photons.

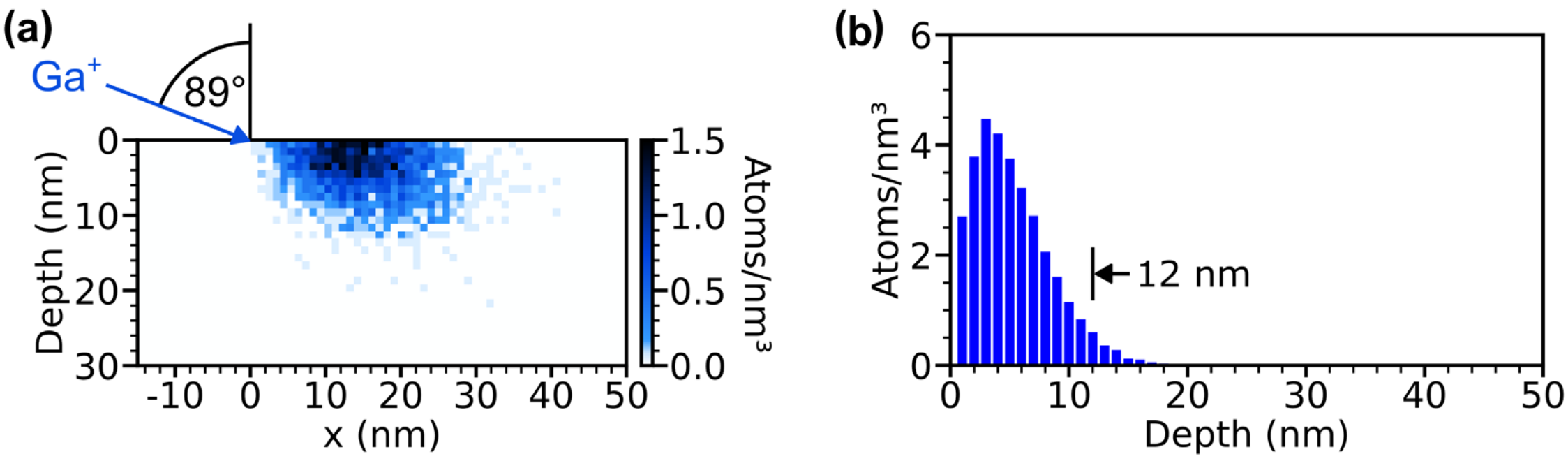

**Fig. S10. Penetration of gallium ions during focused ion beam milling.** (a) Simulated gallium distribution in $VO_2$ at an incidence angle of 89°, obtained by the TRIM software, a package of SRIM (*66*). The absolute scale in atoms/$nm^3$ is obtained from knowledge of the ion beam parameters and illumination time. (b) Simulated depth profile of Ga. We see that the general implantation is low and restricted to the surface; 95 percent of the incident gallium ions are absorbed within ~11.5 nm which is low compared to the ~200 nm of probed material in the electron diffraction experiments.

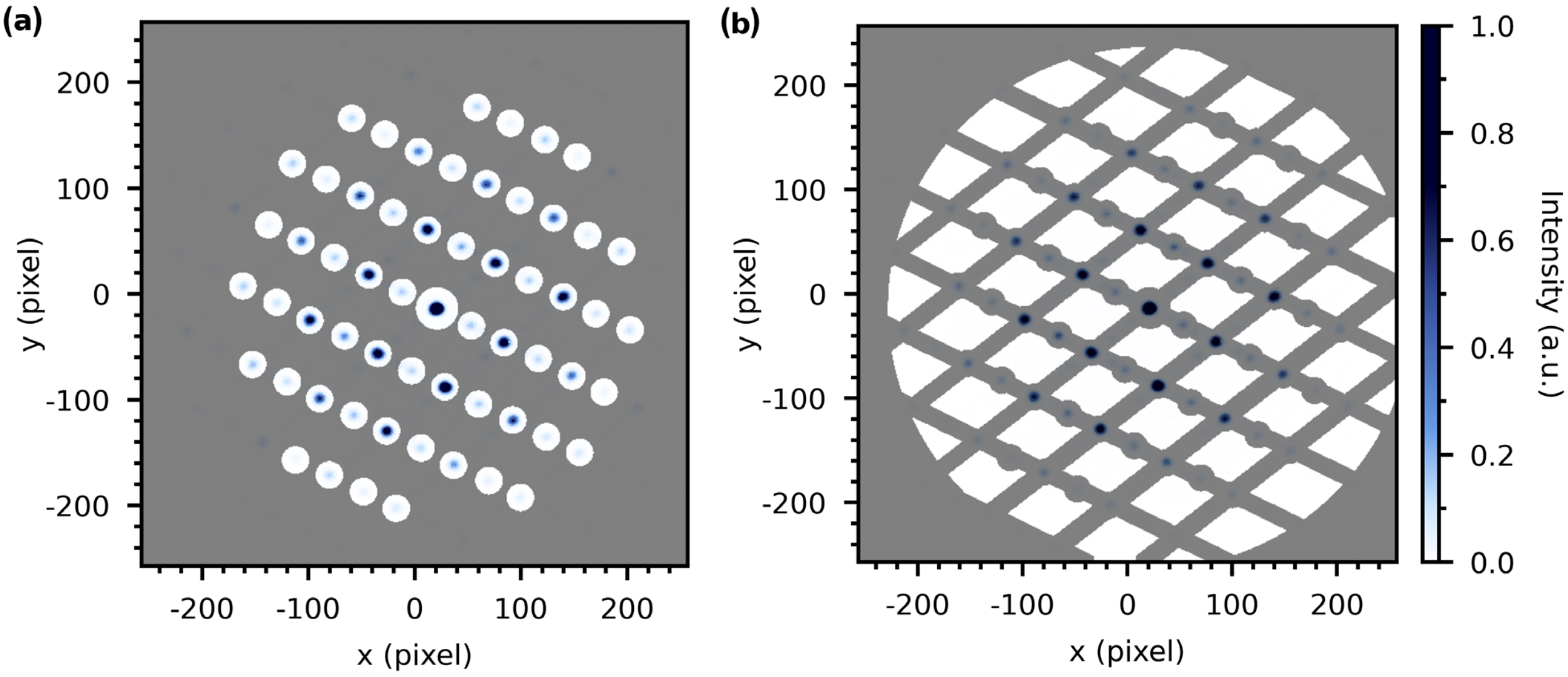

**Fig. S11. Masks for extraction of diffraction intensities.** (a) Mask for the rutile and monoclinic diffraction spots. The direct beam (large circle in the center) is excluded from the analysis. We restrict analysis to peaks below a reciprocal distance of 12 $nm^{-1}$ from the direct beam. (b) Mask used for the diffuse scattering intensity. Supplementary Fig. S12 shows a comparison to an analysis that includes semi-diffuse data between the Bragg spots.

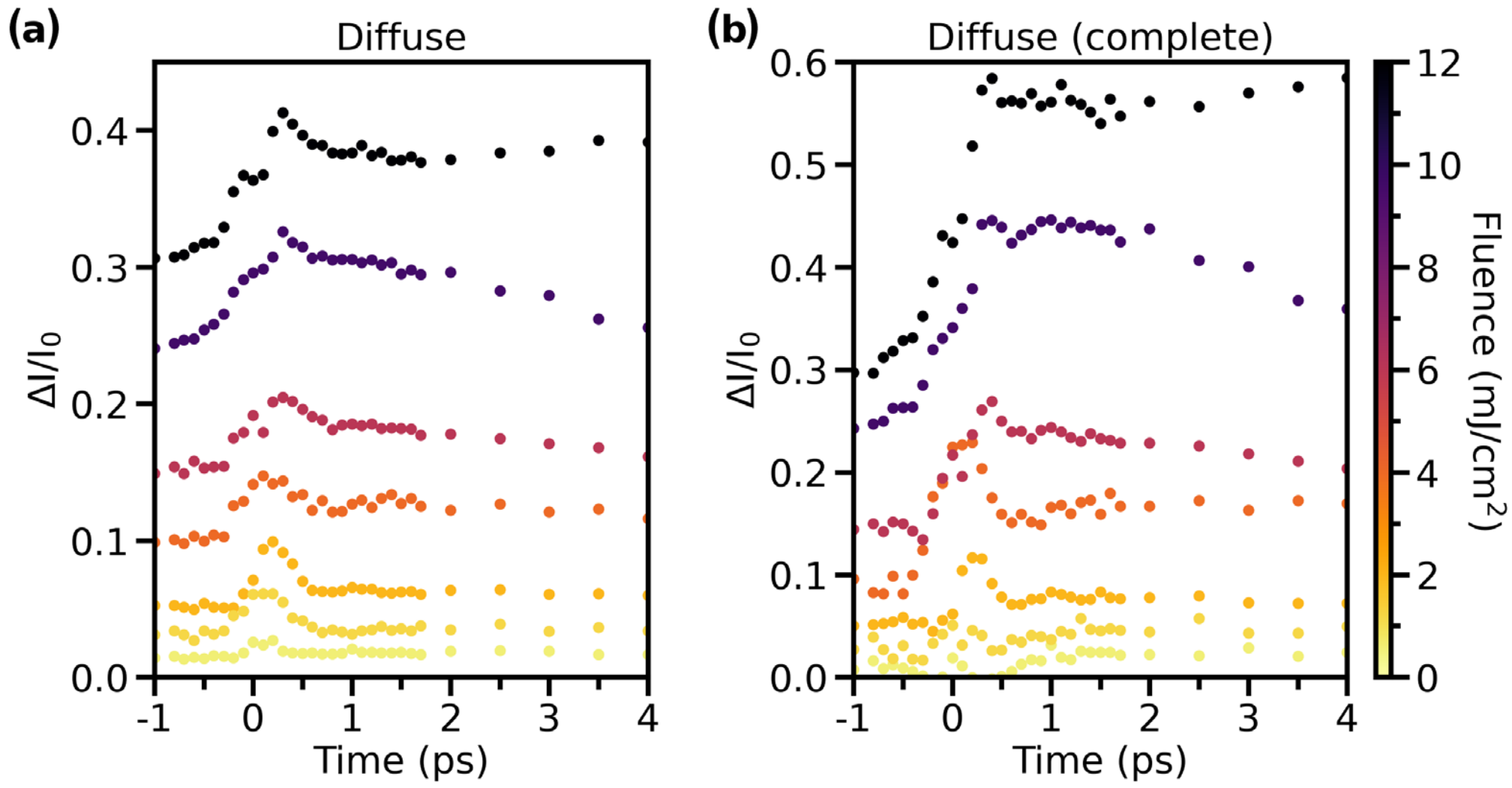

**Fig. S12. Independence of the recorded diffuse dynamics on the analyzed area.** (a) Diffuse dynamics as presented in Figs. 2-3 using the masks depicted in Suppl. Fig. S11b. (b) Diffuse plus semi-diffuse dynamics in the full area without monoclinic or rutile diffraction spots using the negative of Suppl. Fig. S11a as mask. We observe qualitatively similar dynamics in both analyses, with a slightly different absolute amplitude due to the different size of the analyzed area.

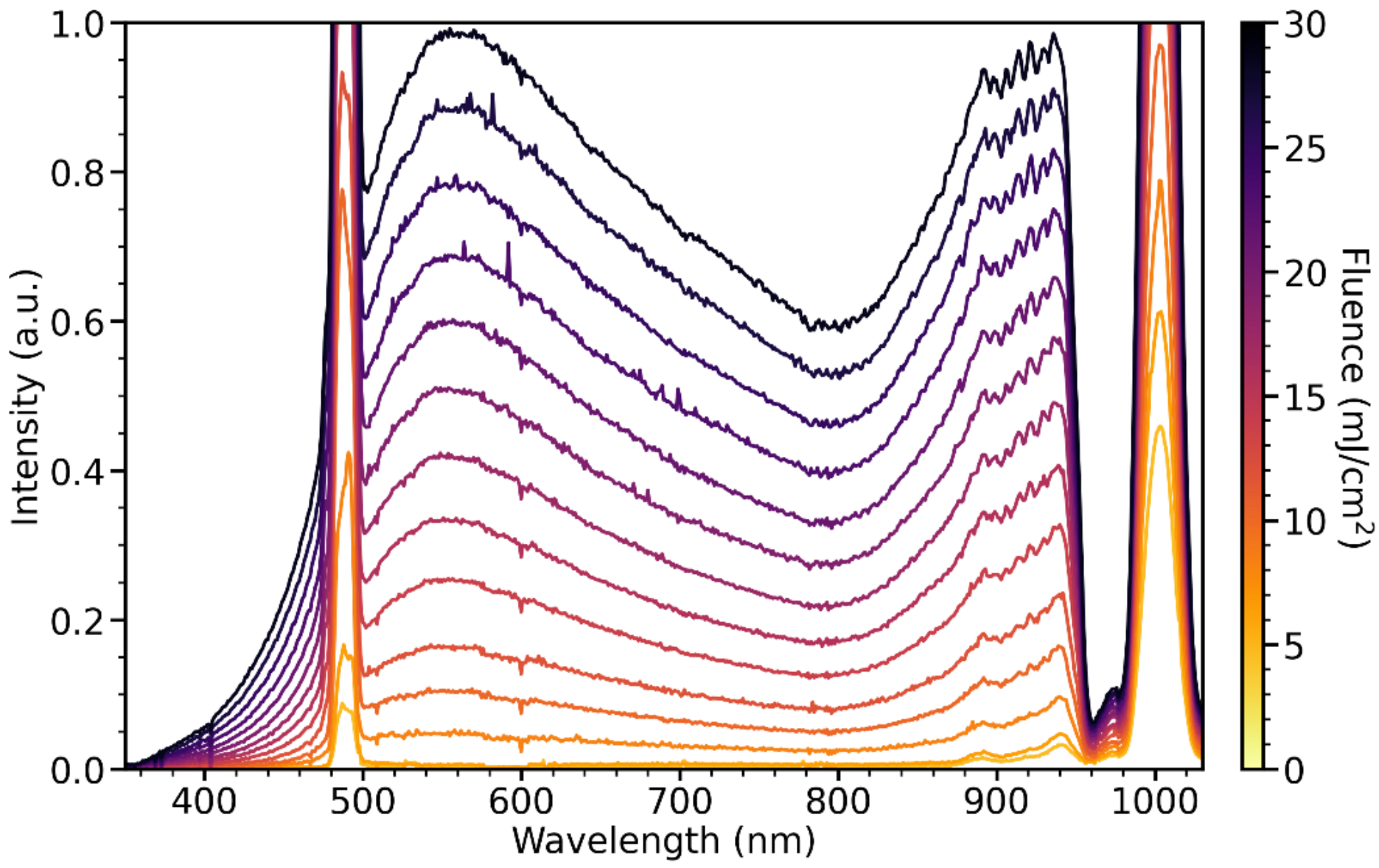

**Fig. S13. Raw thermal emission spectra at different excitation fluences.** The spectrum does not shift significantly when changing the excitation fluence by a factor of thirty (black to yellow). The recorded relative spectral shape is a product of the spectrometer's efficiency curve and the real emission spectrum.

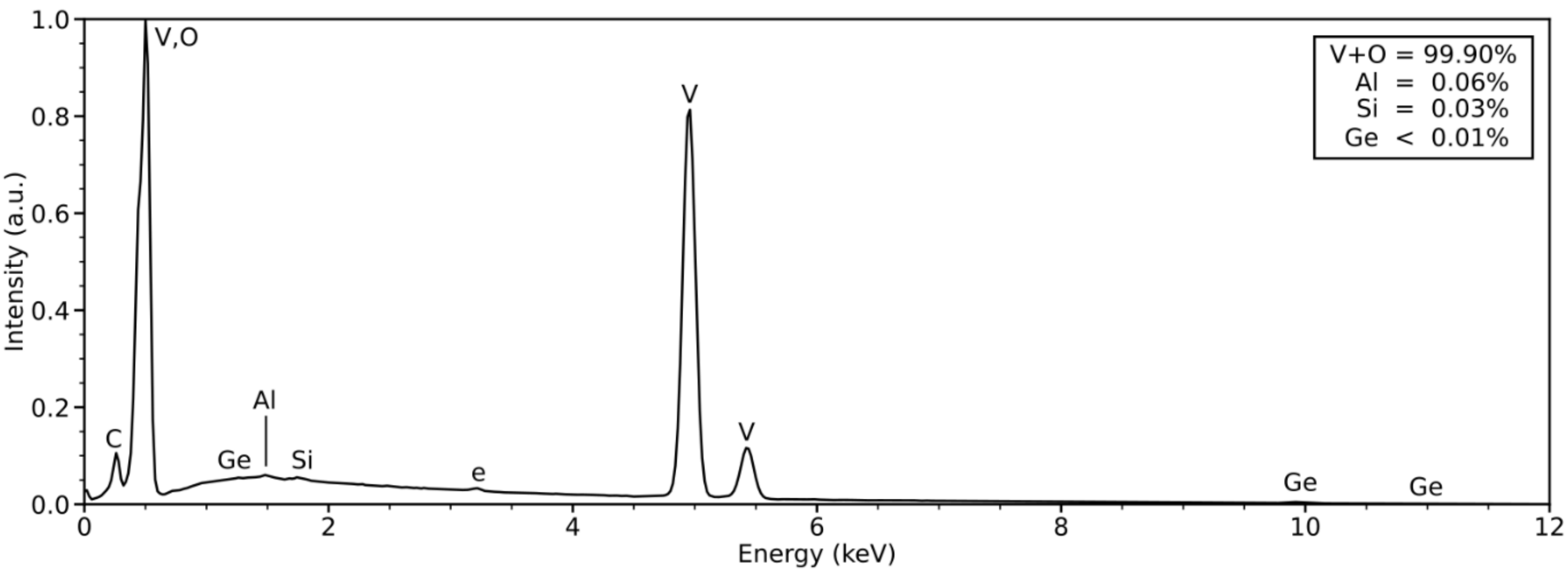

**Fig. S14. Energy-dispersive X-ray spectroscopy.** Besides vanadium and oxygen, we see less than 0.1% of Al, Si and Ge, and no other chemical elements. The small peak at ~3.2 keV is an escape peak of the strong vanadium feature at ~4.9 keV. The broad hump at 1-4 keV is Bremsstrahlung. The peak at ~0.25 keV is from atmospheric carbon in the detection chamber and present in any such experiment, therefore excluded from the analysis.

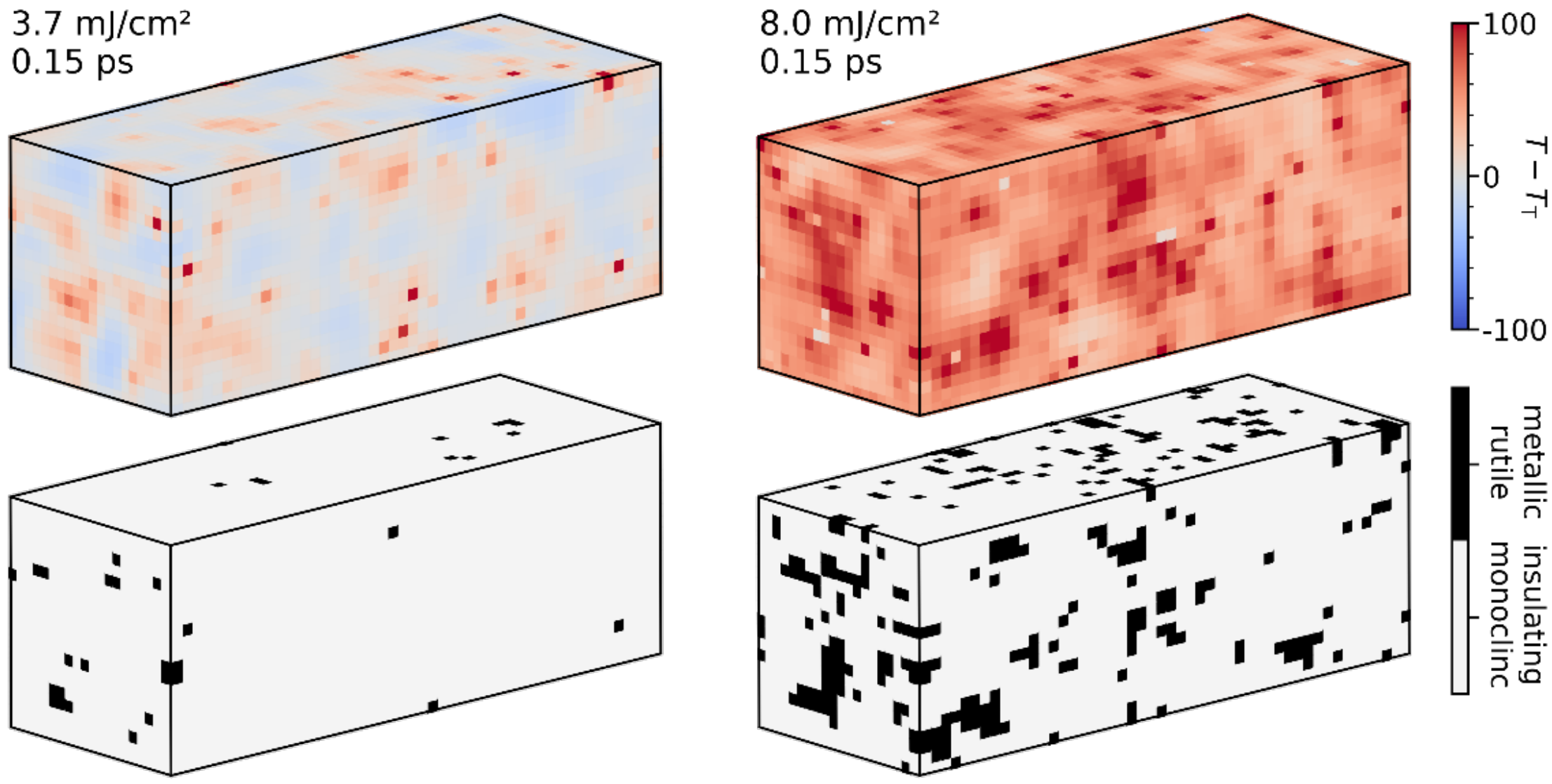

**Movie 1.** Simulated temperature and phase distribution during and after laser incidence. Left column, dynamics at an excitation fluence of 3.7 mJ/cm$^2$. Right column, dynamics at an excitation fluence of 8 mJ/cm$^2$. The upper panels show temperature from cold (blue) to warm (red) and the lower panels show monoclinic/insulating phase (white) and rutile/metallic phase (black) as a function of time.